\documentclass[prb,twocolumn]{revtex4}
\usepackage{graphicx}
\usepackage{hyperref}
\usepackage{amssymb}

\usepackage{float}

\usepackage{booktabs}
\usepackage{amsmath} % or simply amstext
\usepackage[utf8]{inputenc}
\usepackage{lmodern}
\usepackage{color}
\definecolor{red}{rgb}{1.0,0.0,0.0}
\usepackage{makecell}
\usepackage{marginnote}
\usepackage{lipsum}

\DeclareMathAlphabet{\bi}{OML}{cmm}{b}{it}

\def\ba{\begin{aligned}}
\def\ea{\end{aligned}}
\def\be{\begin{equation}}
\def\ee{\end{equation}}
\def\bearr{\begin{eqnarray}}
\def\eearr{\end{eqnarray}}

\usepackage{multirow}
\usepackage{tikz}

\begin{document}
\title{Nonlinear magnetotransport in a two-dimensional system with merging Dirac points}
\bigskip
\author{Ojasvi Pal$^1$, Bashab Dey$^{1,2}$ and Tarun Kanti Ghosh$^1$\\
\normalsize
$^1$Department of Physics, Indian Institute of Technology-Kanpur,
Kanpur-208016, India\\
$^2$Institute for Theoretical Physics, University of Regensburg, Regensburg-93053, Germany}
%\date{\today}
\begin{abstract}
We study the linear, second-order nonlinear (NL) current and voltage responses of a two-dimensional gapped semi-Dirac system with merging Dirac nodes along the $x$ direction under the influence of a weak magnetic field ($B$), using the semiclassical Boltzmann formalism. We investigate the effect of band geometric quantities like Berry curvature and orbital magnetic moment (OMM) in the responses up to linear order in $B$. We derive exact analytical expressions of the linear magnetoconductivities, second-harmonic NL anomalous Hall (NAH), and anomalous velocity and Lorentz force induced (NAL) conductivities, unveiling their dependence on Fermi energy and a gap parameter $\delta_0$. For $\delta_0 > 0$, the Fermi surface topology changes at a particular Fermi energy, which is reflected in the nature of conductivities through a kink. The ratio of NAL and NAH conductivities is found to be independent of $\delta_0$ and inversely related to Fermi energy. The NL dc current exhibits distinct orientations depending on the Fermi energy, magnetic field, polarization of the electromagnetic wave. In the presence of magnetic field, the NL dc current vector can be rotated through large angles on variation of Fermi energy. For high Fermi energies, the NL dc current is directed nearly along the $y$-axis for $x$-polarized and low-frequency circularly polarized light, whereas it aligns close to $x$-axis for high-frequency circularly polarized light. These orientations of the NL dc current are predominantly governed by the mirror symmetry of the system along the $x$ direction. Additionally, we also study the NL voltage responses of the system by applying current along the $x$ and $y$ directions. The system exhibits asymmetry in the $B$-dependencies of the NL resistivities for the two current directions, which may serve as an experimentally relevant signature for band geometric quantities and merging Dirac nodes in such systems.
\end{abstract}
\maketitle
\section{Introduction}
   The exfoliation of graphene in 2004 marked an important epoch in the history of 2D materials due to its unique linear-Dirac spectrum and exotic transport properties\cite{graphene1,graphene2}. The two subbands of graphene touch at two inequivalent points in the reciprocal space named Dirac points, effectively describing the low-energy properties. These Dirac points can be manipulated by varying band parameters such as interaction strength and hopping amplitudes, resulting in the motion of Dirac points. Other two-dimensional physical systems with such spectrum have been observed in organic conductor $\alpha$-(BEDT-TTF)$_2$I$_3$ under pressure\cite{organic1,organic2,organic3,organic4},  8-$Pmmn$ borophene\cite{borophene1,borophene2,borophene3}, artificially fabricated nanostructures\cite{nano5,nano1,nano2,nano3,nano4} and ultracold atoms\cite{cold1,cold2}.
   
   Several studies discovered the merging of Dirac points in the electronic spectrum of two-dimensional systems\cite{merging1,merging2}. The merging of a pair of Dirac points into a single one shows the existence of a topological Lifshitz transition which marks the separation between the semimetallic phase with two disconnected Fermi surfaces and an insulating gapped phase. This also leads to special semi-Dirac dispersion hosting massive fermion behavior along one direction while massless Dirac characteristics in the orthogonal direction\cite{merging3,merging4}. It has been predicted that materials like TiO$_2/$VO$_2$ nanostructures under quantum confinement\cite{tio2} and dielectric photonic crystals\cite{photonic} can exhibit such low-energy dispersions. The merging of Dirac points has been observed experimentally in optical lattices\cite{lattice}, microwave cavities\cite{nano3} and recently, in potassium deposited few-layer black phosphorous \cite{black}. The transport properties such as diffusion\cite{diffusion}, optical conductivity\cite{opto1,opto2,opto3,opto4}, formation of Landau levels spectra under magnetic field\cite{landau1,landau2},  magneto-optical conductivity\cite{moptical}, dynamic polarization and plasmons\cite{tapash} has been studied extensively for semi-Dirac systems. The Landau levels and transport properties for a semi-Dirac nanoribbon were discussed in recent work\cite{basu2}. Very recent studies probed the topological phases of a Chern insulator in such systems by tuning the strength of a circularly polarized light\cite{kush2} and in the presence of extended range hopping\cite{basu}.
   
   The topological behavior of the bands is manifested in the Berry curvature and OMM of the electrons which can significantly affect the linear and the NL transport properties\cite{xiao,moore2}. Some well-known examples in the linear response regime are anomalous Hall effect\cite{anomalous1,anomalous2,anomalous3}, anomalous thermal Hall effect\cite{thermal1,thermal2}, planar Hall effect\cite{phe1,phe2}, magnetoresistance\cite{mr1,mr2}. The discovery of the NL anomalous Hall effect induced by the Berry curvature dipole\cite{fu} in the time-reversal symmetric (TRS) system accelerated the investigation of other NL transport phenomena\cite{nl1,nl2,nl3,nl4,nl5,nl6,nl7,nl8,nl9}. Moreover, it has been realized that such NL transport responses in 2D Dirac systems survive either in the presence of spin-orbit coupling which results in tilting of the Dirac cone, or higher order warping of the Fermi surface\cite{fu, warping}. Recent studies reveal that the low-energy  Hamiltonian that features a pair of Dirac points separated by a saddle point or the merging of two Dirac points can give rise to Berry curvature dipole induced NL transport properties\cite{kush}. However, the linear and NL current responses of such systems in the presence of a magnetic field are not explored. 
   
   In this work, we calculate the contribution of Berry curvature and OMM to the linear and NL conductivities in the presence of a weak magnetic field, using the Boltzmann approach. We also study the second-order NL magnetoresistivity of the system for two different orientations of the applied current. The anisotropy in the nature of $B$-dependencies of the NL resistivities may act as an experimental probe for band geometric quantities as well as merging Dirac nodes in these systems.
          
  This paper is structured as follows: In Sec. \ref{II}, we present the general formulas to calculate the second-order NL current responses in the presence of a weak magnetic field. In particular, we discuss the contribution of Berry curvature and OMM to the NL magnetoconductivities. In Sec. \ref{III}, we provide a discussion on the 2D gapped semi-Dirac model with merging Dirac nodes. In Sec. \ref{IV}, we present the results of linear and NL magnetoconductivities. We further analyze our results and discuss the dependence of all the contributions on Fermi energy and other system parameters in its subsequent subsections. We also discuss the orientations of NL  DC current in response to the linearly and circularly polarized light. Section \ref{V} is dedicated to the discussion of the second-order NL voltage responses. Finally, we summarize our main results in Sec. \ref{VI}.
\section{Theoretical Formulation}\label{II}
In this section, we provide the general formalism to calculate the second-order NL current responses in presence of the electric field which oscillates in time but uniform in space, $ \textbf{E}(t)=\text{Re}[\textbf{E}e^{-i{\omega}t}]=({1}/{2})[\textbf{E}e^{-i{\omega}t}+\mathbf{E}^{\ast}e^{i{\omega}t}] $ with $\textbf{E}={E_x}\hat{\textbf{x}}+{E_y}\hat{\textbf{y}}$, where ${E_x},{E_y}\in\mathbb{C}$ and a static magnetic field $\mathbf{B}$. Theoretically, for an applied electric field $ \textbf{E}(t)$, the linear current of fundamental frequency $j_{a}^{\omega}=\sigma_{ab}E_{b}$ and the NL current $j_{a}^{\text{NL}}=\text{Re}[j_{a}^{(0)}+j_{a}^{(2\omega)}e^{-2i\omega{t}}]$  are measured. Here, $j_{a}^{(0)}=\chi_{abc}^{(0)}E_{b}E_{c}^{*}$ describes the NL DC current and $j_{a}^{(2\omega)}=\chi_{abc}^{(2\omega)}E_{b}E_{c}$ describes the second-harmonic (SH) current and the subscript $a$, $b$ and $c$ are the coordinate indices. 

  The charge current is defined as ${\mathbf{j}}(t) = -e\int[d\mathbf{k}]{D}_{\mathbf{k}}\dot{\mathbf{r}}{f}(t)$, where $[d\mathbf{k}]={d^2}k/{(2\pi)}^2$, $f(t)$ denotes the nonequilibrium distribution function (NDF), ${D}_{\mathbf{k}} = [1+({e}/{\hbar})(\mathbf{B} \cdot \boldsymbol{\Omega})]$ is the phase-space modifying factor\cite{xiao} with  $\boldsymbol{\Omega}$ as the Berry curvature. For simplicity, hereafter we will denote ${D}_{\mathbf{k}}$ by $D$, omitting the implied $k$ dependence. 
  The modified semiclassical equations of motion (including Berry curvature and OMM) for the configuration ($\mathbf{E}\perp\mathbf{B}$) are given by\cite{eom1,xiao,moore}
\begin{align}
{\label{EOM1}}
\mathbf{\dot r}&=\frac{1}{{D}}\bigg[{\tilde{\boldsymbol{v}}}_{\mathbf{k}}+\frac{e}{\hbar}(\mathbf{E}(t) \times  \boldsymbol{\Omega} )\bigg],\\
{\label{EOM2}}
\hbar \mathbf{\dot k} &= \frac{1}{{D}}\bigg[-e \mathbf{E}(t)- e(\tilde{\boldsymbol{v}}_{\mathbf{k}} \times \mathbf{B})\bigg].
\end{align}
The semiclassical band velocity is defined as $\hbar\tilde{\boldsymbol{v}}_{\mathbf{k}} =  {\nabla}_{\mathbf{k}}{\tilde{\epsilon}}_{\mathbf{k}}$, where ${\tilde{\epsilon}}_{\mathbf{k}}$ = ${{\epsilon}}_{\mathbf{k}}- {{\epsilon}}_{\mathbf{k}}^{m}$ is the modified band energy due to Zeeman-like coupling of OMM with the external magnetic field. The OMM modified velocity can be expressed as $\tilde{\boldsymbol{v}}_{\mathbf{k}} = {\boldsymbol{v}}_{\mathbf{k}} -{\boldsymbol{v}}_{\mathbf{k}}^{{m}}$ with $\hbar{\boldsymbol{v}}_{\mathbf{k}}^{{m}}={\nabla}_{\mathbf{k}}(\mathbf{m}\cdot \mathbf{B})$. The Berry curvature for the $n$-th band can be computed using
$\boldsymbol{\Omega}^{n}=-\text{Im}[\langle{\nabla}_{\mathbf{k}}{u}^n_{\mathbf{k}}\vert\times\vert{\nabla}_{\mathbf{k}}u^n_{\mathbf{k}}\rangle]$, where $\vert{u}^n_{\mathbf{k}}\rangle$ is the unperturbed  eigenstate\cite{BC1,xiao}. The OMM, generated by the semiclassical self-rotation of the Bloch wave packet, can be evaluated using $\mathbf{m}^{n}=-({e}/{2\hbar})\textnormal{Im}[\langle{\nabla}_{\mathbf{k}}{u}^n_{\mathbf{k}}\vert\times(H-\epsilon_{\mathbf{k}}^{n})\vert{\nabla}_{\mathbf{k}}u^n_{\mathbf{k}}\rangle]$\cite{BC2}. The Boltzmann transport equation within the relaxation time approximation to obtain the non-equilibrium distribution function (NDF) $f(t)$ is given by\cite{ash}
\begin{equation}\label{BTE}
\frac{\partial{f(t)}}{{\partial}t}+ \dot{\mathbf{k}} \cdot\boldsymbol{\nabla}_{\mathbf{k}}f(t) =-\frac{f(t)-\tilde{f}_{\textnormal{eq}}}{\tau}.
\end{equation}
Here, ${\tilde{f}}_{\textnormal{eq}} = {[1+ e^{\beta(\tilde{\epsilon}_{\mathbf{k}}-\mu)}]}^{-1}$ is the Fermi-Dirac distribution function and $\tau$ is the relaxation time which is considered constant (energy independent) in our case. The NDF can be expressed as $ f(t)=\tilde{f}_{\textnormal{eq}}+\sum_{n=1}^{\infty}f_n(t) $, where the non-equilibrium part of the NDF  can be understood as a power series of the electric field i.e., $f_n\propto {E^n}$. The recursive equation of $f_n$ can be obtained from Eq. (\ref{BTE}) to get the NDF up to quadratic order in an electric field. The general expressions of the linear response current are discussed in details in  Appendix \ref{a} and the corresponding nonzero linear conductivities are given by Eqs. (\ref{drude}), (\ref{lorentz})-(\ref{omm}). These linear conductivities do not have an explicit role in the NL conductivities, but are used to calculate the NL resistivities, which will be discussed in a later section.
\subsection*{ Second-order nonlinear current responses}
  To calculate the second-order NL current responses quadratic in $E$ and up to linear order in $B$, we consider the ansatz for NDF quadratic in $E$,
\begin{equation}
{f_{2}}(t)=f_{2}^{0}+f_{2}^{0 {\ast}}+f_{2}^{2\omega}e^{-i{2\omega}t}+f_{2}^{2\omega {\ast}}e^{i{2\omega}t},
\end{equation}
where $f_{2}^{0}$ denotes the rectification (dc) part and $f_{2}^{2\omega}$ denotes the SH part of the NDF. Substituting it in Eq. (\ref{BTE})
% and equating the coefficients proportional to $e^{-i{2\omega}t}$, we obtain
%\begin{equation}
%-2i\omega f_{2}^{2\omega}-\frac{e}{\hbar D}\Big[\frac{1}{2}(\mathbf{E}\cdot\boldsymbol{\nabla}_{\mathbf{k}}{f}_{1}^{\omega})+(\tilde{\boldsymbol{v}}_{\mathbf{k}} \times \mathbf{B})\cdot\boldsymbol{\nabla}_{\mathbf{k}}f_{2}^{2\omega}\Big]=-\frac{ f_{2}^{2\omega}}{\tau},
%\end{equation}
and applying the Zener-Jones method\cite{jones1}, we can express  $f_{2}^{2\omega}$ in terms of the infinite series of Lorentz force operator $\hat{L}_{B}=({e}/{\hbar})[(\tilde{\boldsymbol{v}}_{\mathbf{k}} \times \mathbf{B})\cdot\boldsymbol{\nabla}_{\mathbf{k}}]$ as\cite{jones2}
\begin{equation}\label{f2w}
f_{2}^{2\omega}=\frac{1}{2}\sum_{\eta=0}^{\infty}\left(\frac{{\tau_{2\omega}}{\hat{L}_{B}}}{D}\right)^{\eta}\left(\frac{e\tau_{2\omega}}{\hbar{D}}\mathbf{E}\cdot\boldsymbol{\nabla}_{\mathbf{k}}{f}_{1}^{\omega}\right),
\end{equation}
where $\tau_{2\omega}=\tau/(1-2i\omega\tau)$ and $f_{2}^{0}= f_{2}^{2\omega}(\mathbf{E}\rightarrow \mathbf{E}^{\ast}, \tau_{2\omega} \rightarrow\tau)$. Here, $f_{1}^{\omega}$ is the first-order correction to the distribution function which is presented in Appendix \ref{a}. Taking into account the weak $B$-field strength, we can express Eq. (\ref{f2w}) as a power series of the magnetic field\cite{moore}. The equilibrium part of NDF  $\tilde{f}_{\textnormal{eq}}$ consists of $B$-dependence through OMM-modified energy, thus it can be expanded via Taylor expansion in terms of $B$ as $\tilde{f}_{\textnormal{eq}}={f}_{\textnormal{eq}}-\epsilon_{m}f'_{\textnormal{eq}}$, where ${f'_{\textnormal{eq}}}\equiv{\partial{f_{\textnormal{eq}}}}/{\partial{\epsilon_k}}$ with ${f}_{\textnormal{eq}}$ defined at $B=0$. 
\begin{center}
\textbf{A. Second-harmonic current}
\end{center}
  For an external electric field oscillating at frequency  $\omega$, we study the SH current  generated at twice the excitation frequency, $j_a^{2\omega}=\chi_{abc}^{(2\omega)}E_{b}E_{c}$. The SH current can be written as  $\textbf{j}_{2}{(t)}=\textbf{j}_{20}(t)+\textbf{j}_{21}(t)$, where the first and second subscript represents the order of $E$ and $B$ respectively. The magnetic field independent SH current can be further expressed as $\textbf{j}_{20}(t)=\textbf{j}_{20}^{2\omega}e^{-2i{\omega}t}+\textbf{j}_{20}^{2\omega\ast}e^{2i{\omega}t}$, where we obtain
\begin{equation}\label{j20}
\textbf{j}_{20}^{2\omega}=\frac{-e^3{\tau}_{\omega}}{4\hbar}\int[d\mathbf{k}]\bigg[(\mathbf{E} \times  \boldsymbol{\Omega})+{\tau}_{2\omega}{\boldsymbol{v}}_{\mathbf{k}}(\mathbf{E}\cdot\boldsymbol{\nabla}_{\mathbf{k}})\bigg](\mathbf{E}\cdot {\boldsymbol{v}}_{\mathbf{k}})f'_{\textnormal{eq}}.
\end{equation}
The corresponding nonzero SH conductivity is given by
\begin{equation}\label{NAH}
\chi_{abc}^{\text{(NAH)}}=-\frac{e^3 \tau_{\omega}}{4\hbar} \varepsilon_{abd}\int[d\mathbf{k}]{\Omega_d} {v_c} {f'_{\textnormal{eq}}}.
\end{equation}
Here, $ \varepsilon_{abd} $ is the Levi-Civita symbol and $abd$ $\in$ $xyz$. The above equation denotes the SH anomalous Hall conductivity $\chi_{abc}^{\text{(NAH)}}$ which is proportional to the dipole moment of Berry curvature over occupied states, defined as $\zeta_{bd}=-\int[d\mathbf{k}]({\nabla_{k_b}}{\epsilon_{\mathbf{k}}}) \Omega_d  f'_{\textnormal{eq}}$. It is evident from the expression that Berry curvature dipole moment survives in time-reversal symmetric and inversion symmetry broken system, unlike the linear anomalous Hall conductivity. The second term in Eq. (\ref{j20}) corresponds to SH Drude conductivity (originating from band velocity) calculated as $\chi_{abc}^{\text{(D)}}=- ({e^3 \tau_{\omega} \tau_{2\omega}}/{4\hbar})\int[d\mathbf{k}]v_a \partial_{k_b} v_c f'_{\textnormal{eq}}$. The SH Drude conductivity vanishes if either of the symmetries between TRS and space inversion symmetry is present in contrast to linear Drude conductivity which is always nonzero.
   
  The SH current linearly dependent on the magnetic field can be written as  $\textbf{j}_{21}(t)=\textbf{j}_{21}^{2\omega}e^{-2i{\omega}t}+\textbf{j}_{21}^{2\omega\ast}e^{2i{\omega}t}$, where
\begin{widetext}
 \begin{equation}\label{nl}
\centering
\begin{aligned}
 \textbf{j}_{21}^{2\omega} &=\frac{e^3}{4\hbar}\int[d\mathbf{k}]\Big\{ (\mathbf{E} \times  \boldsymbol{\Omega})\Big[-\tau_\omega^{2} \hat{L}(\mathbf{E}\cdot {\boldsymbol{v}}_{\mathbf{k}})f'_{\textnormal{eq}} \Big]
 +{\tau_\omega}{\tau_{2\omega}}{\boldsymbol{v}}_{\mathbf{k}}\mathbf{E}\cdot \Big[\frac{e}{\hbar}(\boldsymbol{\Omega}\cdot\mathbf{B})\boldsymbol{\nabla}_{\mathbf{k}}(\mathbf{E}\cdot {\boldsymbol{v}}_{\mathbf{k}}f'_{\textnormal{eq}})+\boldsymbol{\nabla}_{\mathbf{k}}\Big(\frac{e}{\hbar}(\boldsymbol{\Omega}\cdot\mathbf{B})(\mathbf{E}\cdot {\boldsymbol{v}}_{\mathbf{k}})f'_{\textnormal{eq}}\Big)\Big]\\
 &+{\tau_\omega}{\tau_{2\omega}}\Big[{\boldsymbol{v}}_{\mathbf{k}}\mathbf{E}\cdot \boldsymbol{\nabla}_{\mathbf{k}}\Big(\mathbf{E}\cdot \big(\boldsymbol{v}_{\mathbf{k}}^{m} f'_{\textnormal{eq}}+\epsilon_{m}\boldsymbol{v}_{\mathbf{k}}f''_{\textnormal{eq}}\big)\Big)+\boldsymbol{v}_{\mathbf{k}}^{m} \mathbf{E}\cdot\boldsymbol{\nabla}_{\mathbf{k}}(\mathbf{E}\cdot {\boldsymbol{v}}_{\mathbf{k}}f'_{\textnormal{eq}})\Big]
\Big\}.
\end{aligned}
\end{equation}
\end{widetext}
The finite contribution for the SH current linear in $B$ is proportional to $ \tau^2$, since terms  $\propto \tau$ and $\tau^3$ vanish due to TRS. It is to be noted that in the presence of SIS but broken  TRS, all these contributions vanish. Therefore breaking of SIS elicit these nonzero SH responses. The SH conductivity in the presence of a magnetic field encompasses three distinct contributions: the combined effects of anomalous velocity and Lorentz force, the OMM, and the Berry curvature correction to the phase-space factor. The SH Hall conductivity emerging from the combined effects of anomalous velocity and Lorentz force $\chi_{abc}^{\text{(NAL)}}$ can be obtained as
\begin{equation}\label{NL-lorentz}
\chi_{abc}^{\text{(NAL)}}=\frac{e^4 \tau_{\omega}^{2}B}{4\hbar^2}\varepsilon_{abd}\int[d\mathbf{k}]\Omega_d (\boldsymbol{v}_{\mathbf{k}}\times\boldsymbol{\nabla}_{\mathbf{k}})_{z}({{v}_c}f'_{\text{eq}}).
\end{equation}
\begin {figure*}
    \centering
    \includegraphics[width=1.0\textwidth]{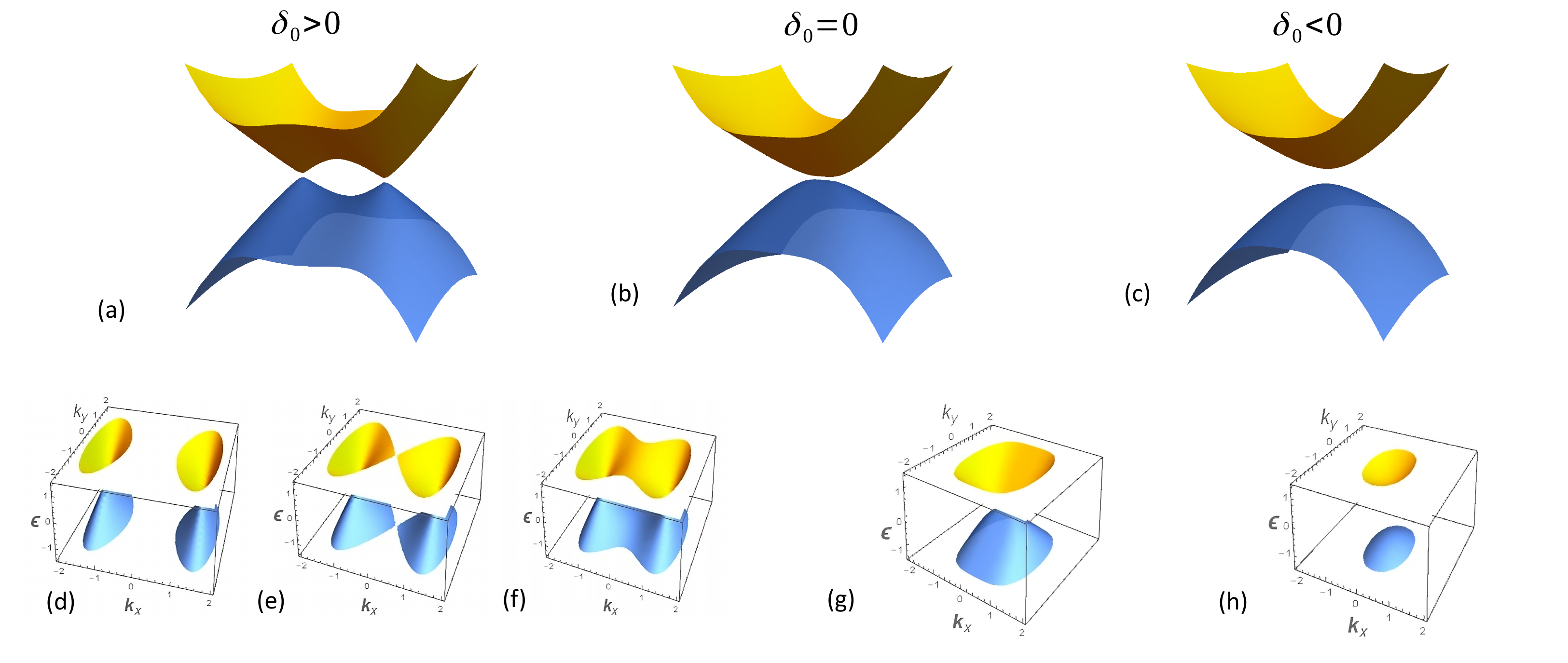}
    \caption{Top panel (a)-(c) represents the energy spectrum of the model given by Eq. (\ref{Hamiltonian}) for different values of gap parameter ${\delta}_{0}$. Plot (a) shows the case of ${\delta}_{0}>0$ describing the phase with two Dirac nodes separated by $2\sqrt{{\delta}_{0}/\alpha}$ distance along $k_x$ direction, (b) ${\delta}_{0}=0$ corresponds to the semi-Dirac form where two Dirac nodes merge and (c) represents the gapped phase. Bottom panel (d)-(h) illustrates the constant Fermi energy contours corresponding to different values of ${\delta}_0$ and Fermi energy. (d)-(f) describes the energy contours existing for three scenarios when ${\delta}_0>0$. (d)represent the two disconnected Fermi surfaces in accordance with the two distinct Dirac nodes till ${\delta}_0>\sqrt{\mu^{2}-{m_{0}^{2}}}$, (e) the two Fermi surfaces get connected by a saddle point at ${\delta}_0=\sqrt{\mu^{2}-{m_{0}^{2}}}$ and (f) the single connected Fermi surface exists as long as ${\delta}_0<\sqrt{\mu^{2}-{m_{0}^{2}}}$. (g) and (h) describes the constant Fermi energy contours for ${\delta}_0=0$ and ${\delta}_0<0$ respectively.  Here, we have used $\beta=10^5$ m/s, $m_0=0.1$ eV, $\alpha$ = 2.7 meV$\cdot$nm$^2$ with m$^*$=13.6 $m_e$ for (TiO$_{2})_{5}$/(VO$_{2})_{3}$, where $m_e$ is the free electron mass.}
\label{fig1}
\end{figure*}
The SH conductivity incited by OMM is given by
\begin{equation}\label{NL-OMM}
\centering
\begin{aligned}
\chi_{abc}^{\text{(OMM)}}&=\frac{e^3 \tau_{\omega}\tau_{2\omega}}{4\hbar}\int[d\mathbf{k}]\big[v_{ma} \partial_{{k}_{b}}({v_c}f'_{\text{eq}})\\
&+v_{a}  \partial_{{k}_{b}}( v_{mc}f'_{\text{eq}}
+ \epsilon_{m}{v_c}f''_{\text{eq}})\big].
\end{aligned}
\end{equation}
Here, $f''_{\text{eq}}$ is the double derivative of $ f_{\text{eq}} $ w.r.t energy. The contribution to the SH conductivity generated by the phase-space factor is obtained as
\begin{equation}\label{NL-B}
\chi_{abc}^{\text{(B)}}=\frac{e^4 \tau_{\omega}\tau_{2\omega}}{4\hbar^2}\int[d\mathbf{k}]v_{a}\big[ (\boldsymbol{\Omega}\cdot\mathbf{B})\partial_{{k}_{b}}+ \partial_{{k}_{b}}(\boldsymbol{\Omega}\cdot\mathbf{B})\big] v_{c}f'_{\text{eq}}.
\end{equation}
We emphasize that all the above three contributions to the SH conductivities in the presence of a magnetic field depend on intrinsic band geometric quantities, namely Berry curvature and OMM. 
\begin{center}
\textbf{B. Nonlinear DC current}
\end{center}
 The second-order response also includes a zero-frequency current known as the photogalvanic effect (PGE). The NL dc current can be expressed as  $j_{a}^{(0)}=\chi_{abc}^{(0)}E_{b}E_{c}^{*}$.
%Unlike the PGE, the response tensors corresponding to second-harmonic current $\chi_{abc}^{(2\omega)}$ and $\chi_{acb}^{(2\omega)}$ are equivalent due to intrinsic permutation symmetry\cite{review}. As a result, PGE depends on the polarization of the electromagnetic wave. 

 The NL dc current arising from the anomalous velocity of Bloch electrons (without magnetic field) can be obtained as\cite{moore2}
%\begin{equation}
%{\mathbf{j}}^{(0)}_{\text{NAH}}= -\frac{e^3\tau}{4\hbar}\int[d\mathbf{k}]\left[(\mathbf{E}\times\boldsymbol{\Omega})\frac{\mathbf{E}^{\ast}\cdot\boldsymbol{v}_\mathbf{k}}{1+i\omega\tau}f'_{\text{eq}}+ \text{c}.\text{c}\right],
%\end{equation}
%For low temperature and low frequency, we obtain
\begin{equation}\label{dc}
\centering
\begin{aligned}
{\mathbf{j}}^{(0)}_{\text{NAH}}&= \frac{1}{1+{\omega^2}{\tau^2}}\Big[2\left({\chi}_{xyy,0}^{\text{(NAH)}}{|E_y|}^{2}\hat{\bf x}+{\chi}_{yxx,0}^{\text{(NAH)}}{|E_x|}^{2}\hat{\bf y} \right)\\
&+\left({\chi}_{xyx,0}^{\text{(NAH)}}{[E_{y}E_x^{*}]}_{+}\hat{\bf x}+{\chi}_{yxy,0}^{\text{(NAH)}}{[E_{y}E_{x}^{*}]}_{+}\hat{\bf y}\right)\\
& - i\omega\tau\left({\chi}_{xyx,0}^{\text{(NAH)}}{[E_{y}E_x^{*}]}_{-}\hat{\bf x}-{\chi}_{yxy,0}^{\text{(NAH)}}{[E_{y}E_x^{*}]}_{-}\hat{\bf y}\right) \Big],
\end{aligned}
\end{equation}
where $ {[E_b E_c^*]}_{\pm}= E_b E_c^*\pm E_c E_b^*$ and $\chi_{abc,0}^{\text{(NAH)}}=\chi_{abc}^{\text{(NAH)}}(\omega=0)$ represents the SH anomalous Hall conductivity given by Eq. (\ref{NAH}) at $\omega=0$. In Eq. (\ref{dc}), the terms in the first parentheses denote the typical photovoltaic effect and the terms in the second (third) parentheses describe the linear (circular) PGE. On the variation of the polarization, the linear PGE (LPGE) is  maximum for linearly polarized light, whereas the circular PGE (CPGE) is  maximum for circularly polarized light. The CPGE current reverses sign when polarization state of the electric field changes from left to right circular. Note that the LPGE vanishes for circularly polarized light, whereas the CPGE  vanishes for linearly polarized light. In Appendix \ref{b}, we present the general formulas for various contributions arising from the geometric quantities to the NL DC current in the presence of a magnetic field. These contributions are expressed in terms of the polarization of the electromagnetic wave.

  In the upcoming sections, we will apply this formalism to the semi-Dirac model with the merging Dirac nodes and  investigate its linear and second-order NL transport properties in the presence of a low $B$-field. 
\section{Gapped semi-Dirac system}\label{III}
The low-energy Hamiltonian that describes the merging of two Dirac nodes has the following form\cite{merging4, diffusion,kush} 
\begin{equation}\label{Hamiltonian}
H(\mathbf{k})=(\alpha{k}_x^2-{\delta}_{0}){\sigma}_{x}+\hbar \beta k_y {\sigma}_{y}+m_0 {\sigma}_{z},
\end{equation}
where  $\boldsymbol{\sigma} = ( \sigma_{x},\sigma_{y},\sigma_{z})$ are the $2\times2$ Pauli matrices in pseudospin space, ${\mathbf{k}}$ is the  crystal momentum having magnitude, $k = \sqrt{{k}_x^2+{k}_y^2}$, ${\delta}_{0}$ and $m_0$ are the gap parameter,  $\alpha={\hbar^2}/{2{m^*}}$ with ${m^\ast}$  as effective mass related to the $x$-direction and $\beta$ is the Dirac velocity along the $y$-direction. In the Hamiltonian given by Eq. (\ref{Hamiltonian}), the mirror symmetry is preserved along the $x$ direction and broken along the $y$ direction. The energy spectrum  is given by
\begin{equation}\label{energy}
\epsilon_{\lambda}({\mathbf{k}})=\lambda \sqrt{(\alpha{k}_x^2-{\delta}_{0})^2+\hbar^2\beta^2 k_y^2 +m_0^2},
\end{equation}
where $\lambda=\pm$ denotes the conduction and valence band respectively. The corresponding band dispersion is shown in Fig. \ref{fig1}. The $x$-component of semiclassical band velocity is calculated as ${\hbar}v_x= 2\alpha {k_x}(\alpha{k_x^2-\delta_{0}})/{\epsilon_{\mathbf{k}}}$  and its $y$-component is $v_y={\hbar}\beta^2  {k_y}/{\epsilon_{\mathbf{k}}}$.
    \begin{figure}[htbp]
\includegraphics[trim={0cm 7.1cm 0.0cm  0cm},clip,width=8.5
cm]{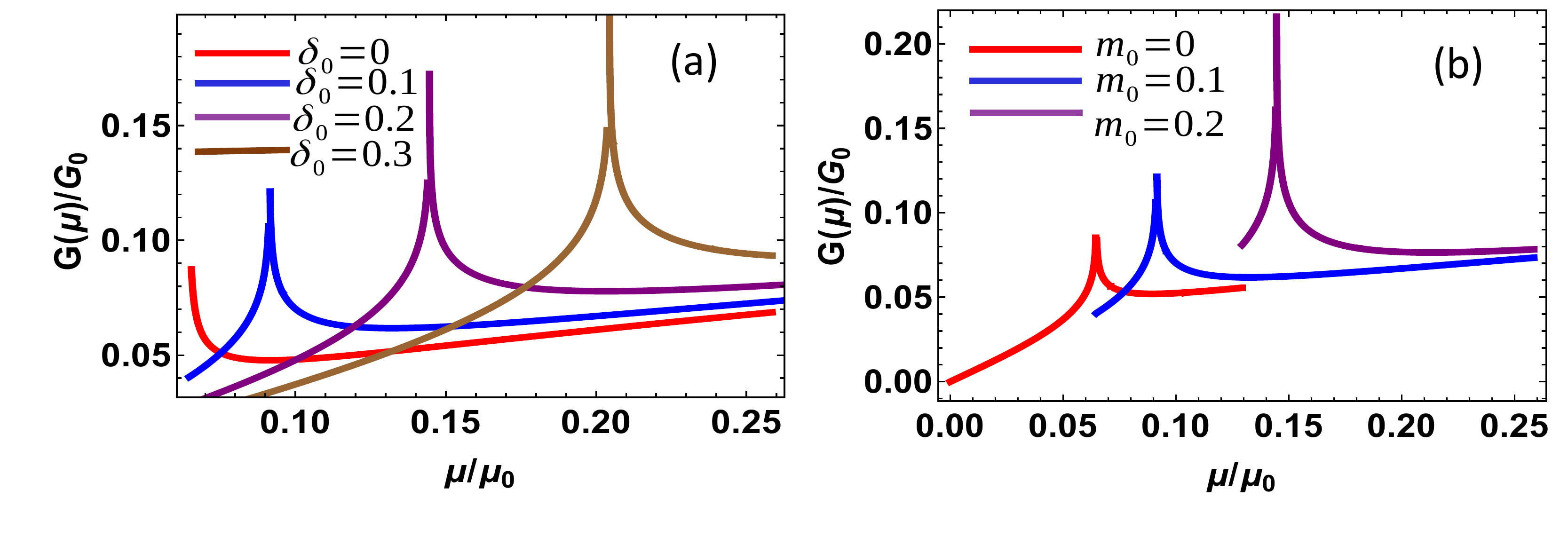}
\caption{Variation of the density of states as a function of Fermi energy: (a) at fixed $m_0=0.1$ eV for different values of $\delta_0$ (in eV) and (b) at fixed $\delta_{0}=0.1$ eV for given values of $m_0$ (in eV). The normalization parameters for density of states and Fermi energy are $G_0=1/\alpha$ and $\mu_0=\hbar^2\beta^2/{\alpha}$ respectively. The parameters used are the same as in Fig. \ref{fig1}.  }  
\label{dos}
\end{figure}
    Equation (\ref{Hamiltonian}) has been termed the ``Universal Hamiltonian" as different types of the spectrum can be obtained by tuning the gap parameter ${\delta}_{0}$\cite{merging4}. The Hamiltonian with ${\delta}_{0}>0$ describes the phase that consists of two Dirac nodes separated by $2\sqrt{{\delta}_{0}/\alpha}$ distance along the $k_x$ axis. In the limit of ${\delta}_{0}=0$, the two Dirac nodes merge and the resulting dispersion exhibits semi-Dirac behavior which is quadratic in the $x$-direction and linear in the $y$-direction. For ${\delta}_{0}<0$, a trivial insulating phase is obtained with a nonzero energy gap. Thus the variation of parameter ${\delta}_{0}$ from negative to positive values drives the transition from an insulating phase to a semi-metallic phase. The mass term $m_0 \sigma_z$ is added in the Hamiltonian to introduce an energy gap at the Dirac nodes. 
 
  We employ the method of parameterization to the constant energy contours $\epsilon{(\mathbf{k})}$  in Eq. (\ref{energy}) considering the sign of $k_x$ (${k_x}\lessgtr0$) in each half-plane. The change of coordinates goes as $\alpha{k}_x^2-{\delta}_{0}=r\cos\phi$, $\hbar\beta{k_y}=r\sin\phi$ and ${\varsigma_k}=sgn(k_x)=\pm$\cite{diffusion}.
The energy spectrum now acquires the simplified form, $\epsilon_{\mathbf{k}}=\pm\sqrt{r^2 +m_0^2 }$ and $\phi$ represents the coordinate along the constant energy contour. The limit of $\phi$ varies according to the topology of the constant energy contours obtained for different energies. The eigenstates are given by
 \begin{equation}
\psi_{{\mathbf{k}}}^{\pm}(\mathbf{r}) = e^{i \mathbf{k} \cdot \mathbf{r}}
 \left (\begin{matrix}
  { \frac{\pm r}{\sqrt{r^2+\big(\sqrt{r^2+m_{0}^{2}}\mp m_0\big)^2}} } \\
{ \frac{\big(\sqrt{r^2+m_{0}^{2}}\mp m_0\big) e^{i \phi}}{\sqrt{r^2+\big(\sqrt{r^2+m_{0}^{2}}\mp m_0\big)^2}}}
\end{matrix} \right).
\end{equation} 
\begin{figure}[htbp]
\includegraphics[trim={0cm 0cm 0cm  0cm},clip,width=8
cm]{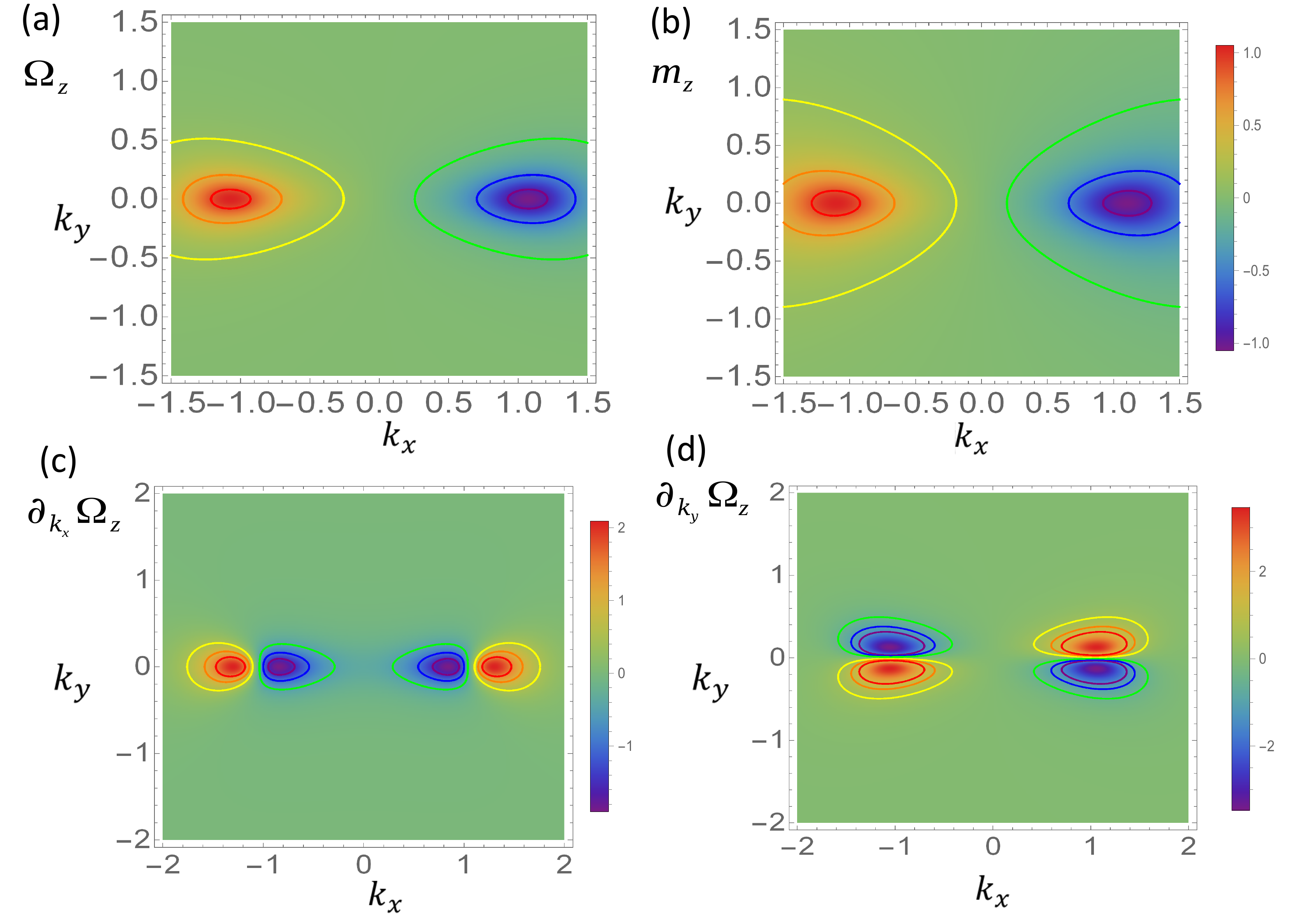}
\caption{Density-contour plot of (a) Berry curvature (in units of $10^{-1}$ nm$^{2}$), (b) OMM (in units of 4 $\alpha$e/$\hbar$), (c) and (d) derivative of Berry curvature with respect to $k_x$ and $k_y$ respectively for the conduction band of the system. Here, $k_x$ and $k_y$ are plotted in units of $\sqrt{{\delta_0}/\alpha}$. The parameters used are the same as in Fig. \ref{fig1}.  }  
\label{contour}
\end{figure}
 Next, we discuss the constant Fermi energy contours corresponding to different values of ${\delta}_0$ and Fermi energy. For a given Fermi energy in the conduction band, the band contains only one minima at ${\delta}_{0}=0$. Hence, we get a single Fermi surface as a result of one nodal point (semi-Dirac node). When ${\delta_0}$ starts to increase, the single minima splits into two minima giving rise to two allowed wave vectors that correspond to two distinct Dirac nodes with linear dispersion. A single connected Fermi surface continues to exist till ${\delta}_0 < \sqrt{\mu^{2}-{m_{0}^{2}}}$ and the area of the Fermi surface gets enhanced due to the presence of extra curvature emerging from the splitting of single minima. These two Dirac points are connected by a saddle point which on a further increase of $\delta_0$ yield two connected Fermi surfaces at  ${\delta}_0 = \sqrt{\mu^{2}-{m_{0}^{2}}}$. The two disconnected Fermi surfaces are formed for ${\delta}_0 > \sqrt{\mu^{2}-{m_{0}^{2}}}$ which marks the onset of a decrease in the Fermi surface area since the bands get narrower with the increase of $\delta_0$. The Fermi surface topology nearly remains uninterrupted with $\delta_0$ for high Fermi energy. Thus the range of $\phi\in [-\phi_{0}, \phi_{0}]$ can be separated into two regions for the case of ${\delta_0} >0$,
 
\begin{equation}\label{limits}
\phi_{0}= \begin{cases} \arccos\left[\frac{-{\delta}_{0}}{\sqrt{\mu^{2}-{m_{0}^{2}}}}\right], & {\delta}_0 < \sqrt{\mu^{2}-{m_{0}^{2}}}, \\ \pi,& {\delta}_0 \geq \sqrt{\mu^{2}-{m_{0}^{2}}}. \end{cases}
\end{equation}
The Jacobian for the transformation of coordinates from ($k_x$, $k_y$) to ($r$, $\phi$) is given by
\begin{equation}
J(r, \phi)=\frac{r}{2 \hbar\beta \sqrt{\alpha(r\cos\phi+\delta_{0})} }.
\end{equation}
We obtain the density of states (DOS) given as
\begin{equation}\label{dos1}
G(\mu)= 2\frac{G_0}{{\pi^2}\sqrt{8}}\sqrt{\frac{{\tilde{\delta}_0^2}+{\tilde{m}_{0}^{2}}{\gamma^2}}{\gamma{\tilde{\delta}_0}}}\begin{cases} K(k), & \gamma<1, \\ k' [K(k')],& \gamma\geq 1, \end{cases}
\end{equation}
where $K(k)$ and $E(k)$ are the complete elliptic integrals of the first and second kind, respectively, with  $k=\sqrt{({1+\gamma})/{2}}$ and $k'=1/k=\sqrt{{2}/{(1+\gamma)}}$ known as the modulus of Jacobian elliptic function and integrals. We introduce $\gamma=\delta_{0}/\sqrt{\mu^{2}-{m_{0}^{2}}}$ as a reduced parameter and  $G_0=1/\alpha$. Note that an overall factor of 2 is multiplied to consider the sign of $k_x$. Here, we define the scaled system parameters as $\tilde{\delta}_0=\delta_0/\mu_0$ and $\tilde{m}_0=m_0/\mu_0$ with $\mu_0=\hbar^2\beta^2/\alpha$. We choose $\mu>m_{0}$ such that the Fermi energy lies above the bulk gap. Expanding the above expression of DOS for  $\gamma \geq 1$ up to leading order in $k'$ gives
\begin{equation}
G(\mu)\simeq \frac{G_0}{32{\pi}}\sqrt{\frac{{\tilde{\delta}_0^2}+{\tilde{m}_{0}^{2}}{\gamma^2}}{\gamma{\tilde{\delta}_0}}}\left(\frac{33+40\gamma+16{\gamma}^{2}}{{(1+\gamma)^{5/2}}}\right).
\end{equation} 
\begin{figure*}
    \centering
    \includegraphics[width=1.0\textwidth]{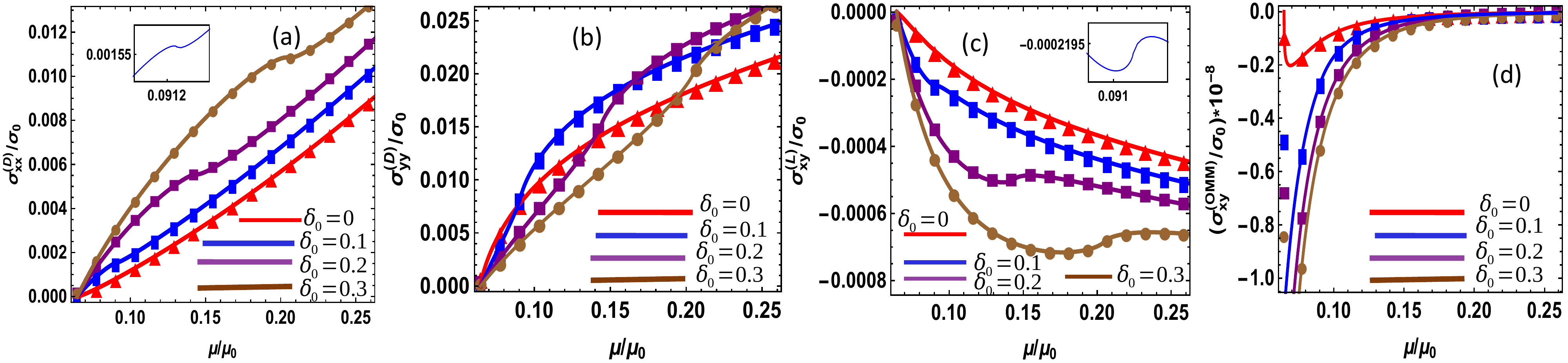}
    \caption{Variation of linear conductivities (up to linear order in $B$) with the Fermi energy for different values of $\delta_0$ (in eV). (a)-(b) Drude conductivity along the $x$ and $y$ direction respectively, (c) Lorentz force induced linear Hall conductivity, and (d) OMM induced linear Hall conductivity. The normalization parameter for linear conductivties  is $\sigma_0=\tau  e^2\beta^2/\alpha$. Here, the solid color curves represent the results of analytical calculations at zero temperature whereas the corresponding colored plot markers are the representatives of the results of numerical calculations performed at temperature $T=34 $ K. The inset is a blow-up of the region around  $\mu=\sqrt{{{\delta}_0^2}+{m_{0}^{2}}}$ for $\delta_0=0.1$ eV. The parameters used are $\beta=10^5$ m/s, $\alpha=2.7$ meV$\cdot$nm$^2$ with $m^*=13.6 m_e$, $\tau=10^{-12}$ s, $B=2$ T and $m_0=0.1$ eV.}
 \label{fig2}
\end{figure*}
Near the band-edge in the limit of $\gamma\rightarrow\infty$, $G(\mu)\propto 1/\sqrt{\delta_0}$. For very low doping, DOS decreases with $\delta_0$ whereas, for high doping, DOS increases with $\delta_0$ and matches with semi-Dirac result in the large $\mu$ limit as shown in Fig. \ref{dos}(a). Note that divergence occurs exactly at the saddle point ($\gamma=1$). Equation \ref{dos1} at $m_0=0$ reduces to the known results\cite{merging4}. We also plotted the variation of DOS with Fermi energy at fixed $\delta_{0}=0.1$ eV for different values of energy gap parameter $m_0$ in Fig. \ref{dos}(b). We find that as we increase the value of $m_0$ from 0.0 to 0.1 eV, the peak of divergence also gets shifted since the allowed range of $\mu$ will also change from $\mu=0.0$ to $0.1$ eV such that $\mu>m_0$ to get the physical results.

  The Berry curvature and OMM of the given Hamiltonian can be calculated as
\begin{equation}
\Omega_{z}=\mp\frac{\hbar\alpha \beta m_{0}k_{x}}{\epsilon_{\mathbf{k}}^{3}}, \hspace{0.5cm}
m_{z}=-\frac{e\alpha \beta m_{0}k_{x}}{\epsilon_{\mathbf{k}}^{2}}.
\end{equation}
It is to be noted that Berry curvature and OMM are zero for the gapless system ($m_0=0$), whereas, in the limit of $\delta_{0}=0$, it remains nonzero. The OMM is the same for both the bands. The density-contour plot of the Berry curvature and OMM for the conduction band are shown in Fig. \ref{contour}. The magnitude of Berry curvature decays rapidly as compared to OMM when Fermi energy shifts away from the band edge.
\section{Results and Discussions}\label{IV}
In this section, we calculate the linear, SH and NL dc current responses of the gapped semi-Dirac system  in the presence of a static magnetic field applied along the $z$ direction. In experimental setups\cite{nl2,translimit1}, the ac frequency lies in the range of $10-1000$ Hz and relaxation time $\tau\sim 10^{-12}$ s, which explains the transport limit, i.e., $\omega\tau\ll 1$. In this limit, $\tau_{\omega}\rightarrow\tau$ and $\tau_{2\omega}\rightarrow\tau$. In our work, we utilize this limit to determine the linear and SH conductivities of the system by substituting $\tau_{\omega}$ and $\tau_{2\omega}$ with $\tau$. In this limit, it is noteworthy that the SH and the NL DC conductivities are nearly equal.

\subsection*{ Linear conductivities}
  We first evaluate the linear conductivities of the system using the general form of the Eqs. (\ref{drude}), (\ref{lorentz})-(\ref{omm}) along with modified coordinates that account for the dispersion anisotropy. In the limit of zero temperature, a derivative of the Fermi–Dirac distribution function is substituted by the Dirac-delta function which allows to perform the integral over energy analytically. 
%The Drude conductivity along the $x$ direction after integration over energy takes the following form
%\begin{equation}
%\sigma_{xx}^{\text{(D)}}=\frac{\sigma_0}{2\pi^2}\frac{{\tilde{\delta}_0^{5/2}}}{\sqrt{\gamma^3({\tilde{\delta}_0^2}+{\tilde{m}_0^2}{\gamma}^{2})}}\int_{{-}\phi_{0}}^{\phi_0}d\phi \cos^{2}\phi {\sqrt{\cos\phi+\gamma}},
%\end{equation}
 Next, we can evaluate the angular integration by considering the appropriate limits of $\phi$ in accordance with the two regimes discussed in Eq. (\ref{limits}) for $\delta_0>0$. Such angular integrals can be expressed in terms of complete elliptic integrals of the first and second kinds. The $xx$-component of Drude conductivity can be calculated using Eq. (\ref{drude}) as
\begin{equation}
\centering
\begin{aligned}\label{exact-xx}
\sigma_{xx}^{\text{(D)}}&=\frac{\sqrt{2}\sigma_0}{15  \pi^2}\frac{{\tilde{\delta}_0^{5/2}}}{\sqrt{\gamma^3({\tilde{\delta}_0^2}+{\tilde{m}_0^2}{\gamma}^{2})}}\\&\begin{cases} \big[(2\gamma^2 +7\gamma-9) K(k)+2(9-2\gamma^2)E(k)\big], 
& \gamma < 1, \\ {2 k}\big[(9-2{\gamma^2}) E(k')+2\gamma({\gamma}-1)K(k')\big],
 & \gamma \geq 1, \end{cases}
\end{aligned}
\end{equation}
where  $\sigma_0=\tau  e^2\beta^2/\alpha$. In view of anisotropic dispersion, $\sigma_{xx}^{\text{(D)}}\neq\sigma_{yy}^{\text{(D)}}$. Following the details of calculation similar to $xx$-term, the $yy$-component of Drude conductivity is calculated to be
\begin{equation}\label{exact-yy}
\centering
\begin{aligned}
\sigma_{yy}^{\text{(D)}}&=\frac{{\sqrt{2}}\sigma_0}{6\pi^2}\frac{\tilde{\delta}_0^{3/2}}{\sqrt{\gamma({\tilde{\delta}_0^2}+{\tilde{m}_0^2}{\gamma}^{2})}}\\&\begin{cases} \big[2\gamma E(k)+(1-\gamma)K(k)\big], & \gamma < 1, \\ {2 k}\big[\gamma E(k')-(\gamma-1)K(k')\big] ,& \gamma \geq 1. \end{cases}
\end{aligned}
\end{equation}
\begin{figure}[htbp]
\includegraphics[trim={0cm 6.6cm 0cm  0cm},clip,width=8.8
cm]{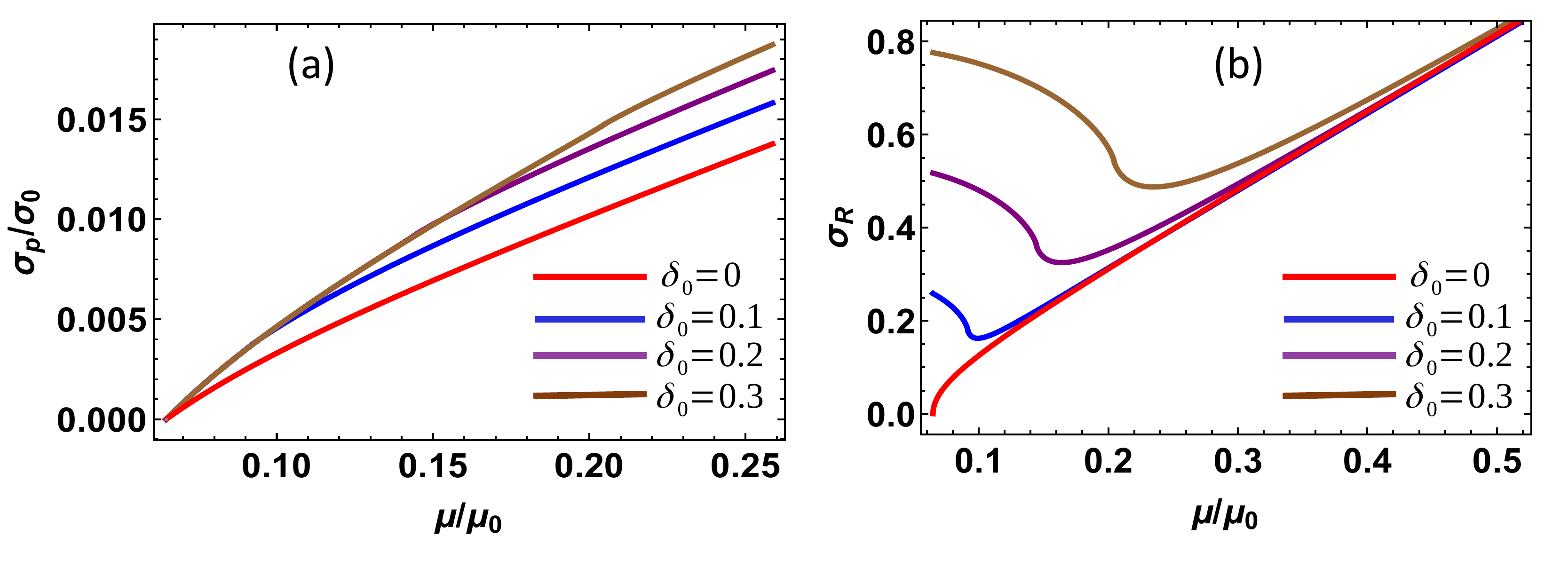}
\caption{(a)-(b) Plots of variation of $\sigma_P\equiv\sqrt{\sigma_{xx}^{\text{(D)}}\sigma_{yy}^{\text{(D)}}}$ and $\sigma_R\equiv\sigma_{xx}^{\text{(D)}}/{\sigma_{yy}^{\text{(D)}}}$ with the Fermi energy for different values of $\delta_0$ (in eV). The solid color curves represent the analytical results calculated at zero temperature. All parameters used are the same as in Fig. \ref{fig2}.}
\label{product-fig}
\end{figure}
We have plotted the Drude conductivities $\sigma_{xx}^{\text{(D)}}$ and $\sigma_{yy}^{\text{(D)}}$ as a function of Fermi energy at the given values of ${\delta_{0}}$ in Figs. \ref{fig2}(a) and \ref{fig2}(b). The behavior of conductivities is expectedly different for $\gamma < 1$ (high Fermi energy with single-connected Fermi surface) and  $\gamma \geq 1$ (low Fermi energy with two Fermi surfaces) due to the particular Fermi surface topology in the two regimes. For the case of $\gamma \geq 1$, expanding the exact analytic expressions given by Eqs. (\ref{exact-xx}) and (\ref{exact-yy}) up to leading order in $k'$, we get
\begin{equation}
\begin{aligned}
\sigma_{xx}^{\text{(D)}}&\simeq\frac{\sigma_0}{16\pi}\frac{{\tilde{\delta}_0^{5/2}}}{\sqrt{\gamma^3({\tilde{\delta}_0^2}+{\tilde{m}_0^2}{\gamma}^{2})}}\left[\frac{(1+2\gamma)(3+4\gamma)}{(\gamma+1)^{3/2}}\right],\\
\sigma_{yy}^{\text{(D)}}&\simeq \frac{\sigma_0}{32\pi}\frac{\tilde{\delta}_0^{3/2}}{\sqrt{\gamma({\tilde{\delta}_0^2}+{\tilde{m}_0^2}{\gamma}^{2})}}\left[\frac{(11+4\gamma)}{(\gamma+1)^{3/2}}\right].
\end{aligned}
\end{equation}
In the limit of $\gamma \rightarrow \infty$ (near the band edge approximation i.e., $\mu\rightarrow m_0$), we find that $\sigma_{xx}^{\text{(D)}}\propto \mu^2 \sqrt{\delta_{0}}$ and $\sigma_{yy}^{\text{(D)}}\propto \mu^2/ \sqrt{\delta_{0}}$. Hence, for low doping, $\sigma_{xx}^{\text{(D)}}$ increases  with both $\mu$ and ${\delta_{0}}$, whereas $\sigma_{yy}^{\text{(D)}}$  increases with the Fermi energy but decreases with $\delta_{0}$. We noticed a small kink in both the Drude conductivities as a consequence of a change in Fermi surface topology exactly at the saddle point ($\gamma=1$) for the fixed positive values of $\delta_0$. For high doping, $\sigma_{xx}^{\text{(D)}}$ and $\sigma_{yy}^{\text{(D)}}$ continues to increase with the Fermi energy. The increasing nature of  Drude conductivities with the Fermi energy can be explained by the monotonous increase of Fermi surface area and velocities with $\mu$ for given $\delta_0$. It is to be noted that for high Fermi energy, the Fermi surface topology relatively remains the same with $\delta_0$.

  We also note the variation for the geometric mean of Drude conductivities defined as $\sigma_P=\sqrt{\sigma_{xx}^{\text{(D)}}\sigma_{yy}^{\text{(D)}}}$ and their ratio $\sigma_R\equiv\sigma_{xx}^{\text{(D)}}/\sigma_{yy}^{\text{(D)}}$ with $\mu$ and $\delta_0$. We are interested in probing the variations of experimentally relevant quantity $\sigma_R$ which is independent of scattering time since the calculation of scattering time may not be straightforward. We find that  $\sigma_P$ is independent  of $\alpha$ and $\beta$. Similar to Drude conductivities, $\sigma_P$ also shows an increase with the Fermi energy. However, the ratio $\sigma_R$ decreases with Fermi energy for low doping while for high doping, it shows an increase with the Fermi energy for a given $\delta_0$. Note that for low doping, the semi-Dirac curve $(\delta_0=0)$ for $\sigma_R$ deviates significantly from the finite $\delta_0$ case. For $\delta_0=0$, $\sigma_R$ increases with the Fermi energy whereas decreases for finite $\delta_0$. For high doping, the resulting ratio curve matches perfectly well with the semi-Dirac case. Next, we discuss the variation of these quantities as a function of $\delta_0$ for a given Fermi energy. We find that for low doping, $\sigma_P$ is nearly constant which is consistent with our results of $\gamma\rightarrow\infty$ limit, while for high doping, $\sigma_P$ shows an increase with $\delta_0$. The ratio $\sigma_R$ increases with $\delta_0$. This variation for the geometric mean and ratio of Drude conductivities are illustrated in Figs. \ref{product-fig}(a) and \ref{product-fig}(b) respectively. 
  
  Now we move to compute the magnetic field-dependent linear conductivities of the system. The Lorentz force-induced Hall conductivity can be calculated using Eq. (\ref{lorentz}) as
\begin{equation}
\begin{aligned}
\sigma_{xy}^{\text{(L)}}&=-\frac{\sqrt{2}\sigma_0\tilde{B}_1}{15 \pi^2 }\frac{\tilde{\delta}_0^{5/2}}{{\sqrt{\gamma}}({{\tilde{\delta}_0^2}+{\tilde{m}_0^2}{\gamma}^{2}})}\\&\begin{cases} \big[(2\gamma^2 +7\gamma-9) K(k)+2(9-2\gamma^2)E(k)\big], 
 &\gamma < 1, \\ {2 k}\big[(9-2\gamma^2) E(k')+2\gamma(\gamma-1)K(k')\big], 
 & \gamma \geq 1. \end{cases}
\end{aligned}
\end{equation}
Here,  $\sigma_{xy}^{\text{(L)}}=-\sigma_{yx}^{\text{(L)}}$ and $\tilde{B}_1=B/B_1$ with $B_1=\hbar^2/(e\tau\alpha)$. Expanding the low-energy expression up to the leading order in $k'$, we find
\begin{equation}
\sigma_{xy}^{\text{(L)}}\simeq\frac{-\sigma_0\tilde{B}_1}{16\pi}\frac{\tilde{\delta}_0^{5/2}}{{\sqrt{\gamma}}({{\tilde{\delta}_0^2}+{\tilde{m}_0^2}{\gamma}^{2}})}
\left[\frac{(1+2\gamma)(3+4\gamma)}{(\gamma+1)^{3/2}}\right].
\end{equation}
In the limit of $\gamma\rightarrow\infty$, it turns out that $\sigma_{xy}^{\text{(L)}} \propto (\mu^2)\sqrt{\delta_{0}}$. The variation of $\sigma_{xy}^{\text{(L)}}$ with the Fermi energy is shown in Fig. \ref{fig2}(c). Below the saddle point for $\gamma > 1$, $\sigma_{xy}^{\text{(L)}}$ increases with the increase in Fermi energy and  ${\delta_{0}}$. In other words, $\sigma_{xy}^{\text{(L)}}$ decreases as Dirac nodes moves closer to each other (${\delta_{0}}\rightarrow0$). As expected, a little kink is observed at the saddle point. Past the saddle point, $\sigma_{xy}^{\text{(L)}}$ continues to monotonically increasing with $\mu$. For high Fermi energy, $\sigma_{xy}^{\text{(L)}}$ increases with $\delta_{0}$. 
  \begin{figure}[htbp]
\includegraphics[trim={0.2cm 0cm 0cm 0cm},clip,width=7.5
cm]{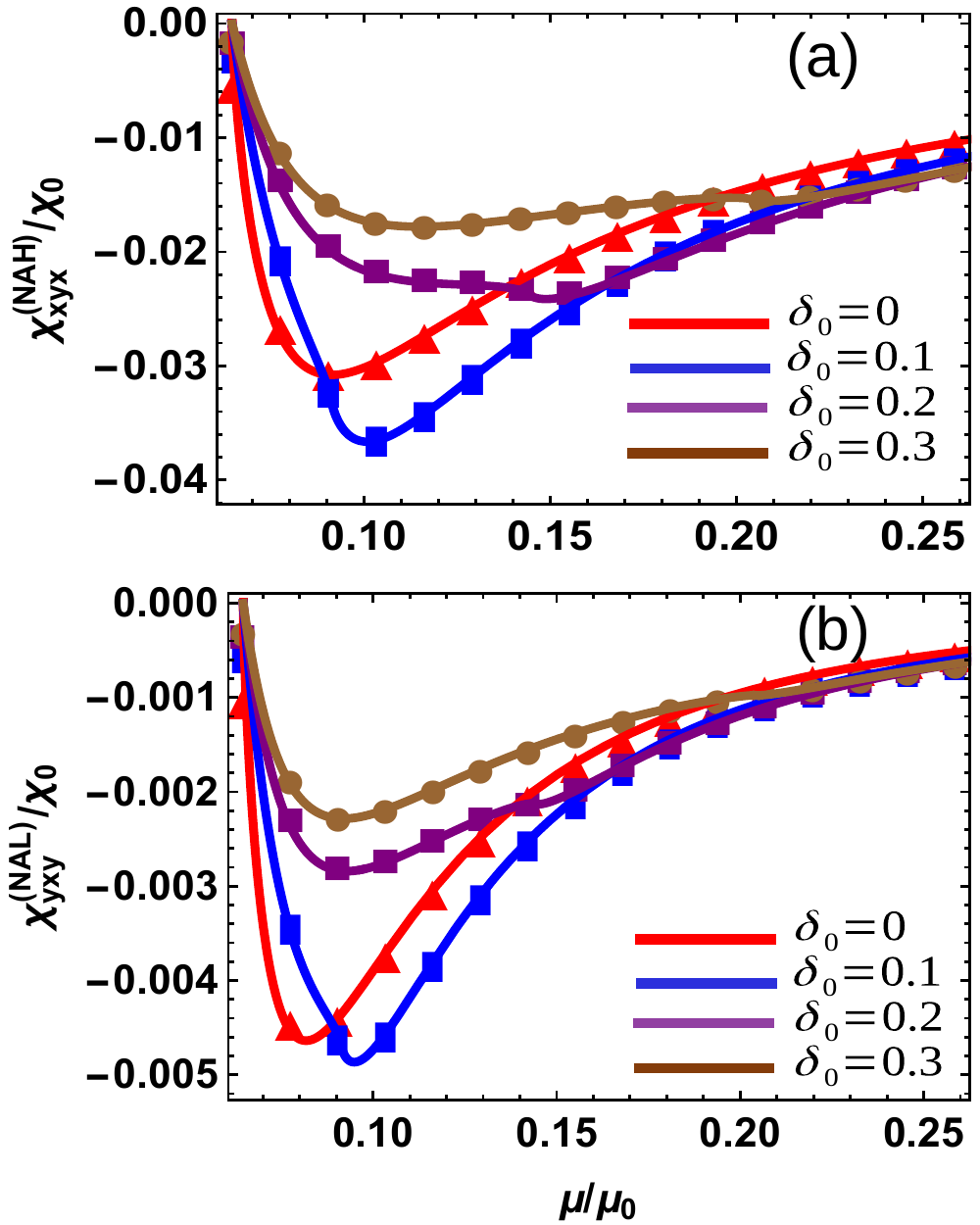}
\caption{(a) Depicts the behavior of SH anomalous Hall conductivity with the Fermi energy. (b) The anomalous velocity and Lorentz force induced SH conductivity as a function of Fermi energy. Both of these plots are plotted for different values of $\delta_0$ in eV. The solid color curves are the results of analytical calculations at zero temperature which matches perfectly well with the numerical results obtained at $T=34$ K (color plot markers). The above SH conductivities are normalized by $\chi_0=e^3\tau\alpha/\hbar^3 \beta$. The parameters used are the same as in Fig. \ref{fig2}. }
\label{fig3}
\end{figure}
  The OMM induced Hall conductivity can be evaluated using Eq. (\ref{omm}) as
\begin{equation}
\centering
\begin{aligned}
\sigma_{xy}^{\text{(OMM)}}=-\sigma_{yx}^{\text{(OMM)}}&=-\frac{\sigma_0 \tilde{m}_0^2\tilde{B}_2}{2\sqrt{2}\pi^2 }\frac{\sqrt{{\gamma^{7}}{\tilde{\delta}_0}}}{({{\tilde{\delta}_0^2}+{\tilde{m}_0^2}{\gamma}^{2}})^2}\\&\begin{cases} \big[2 E(k)-{(1-\gamma)}K(k)\big], & \gamma < 1, \\ {2k}\big[ E(k')\big]  ,& \gamma \geq 1, \end{cases}
\end{aligned}
\end{equation}
where $\tilde{B}_2=B/B_2$ with $B_2=\tau\hbar^4\beta^4/(e\alpha^3)$. Expanding the above expression for $\gamma \geq 1$, we find
\begin{equation}
\sigma_{xy}^{\text{(OMM)}}\simeq\frac{\sigma_0 \tilde{m}_0^2\tilde{B}_2}{64\pi }\frac{\sqrt{{\gamma^{7}}{\tilde{\delta}_0}}}{({{\tilde{\delta}_0^2}+{\tilde{m}_0^2}{\gamma}^{2}})^2}
\left[\frac{(1+4\gamma)(5+4\gamma)}{(\gamma+1)^{3/2}}\right].
\end{equation}
\begin{figure}[htbp]
\includegraphics[trim={0.0cm 0cm 0cm 0cm},clip,width=7.8
cm]{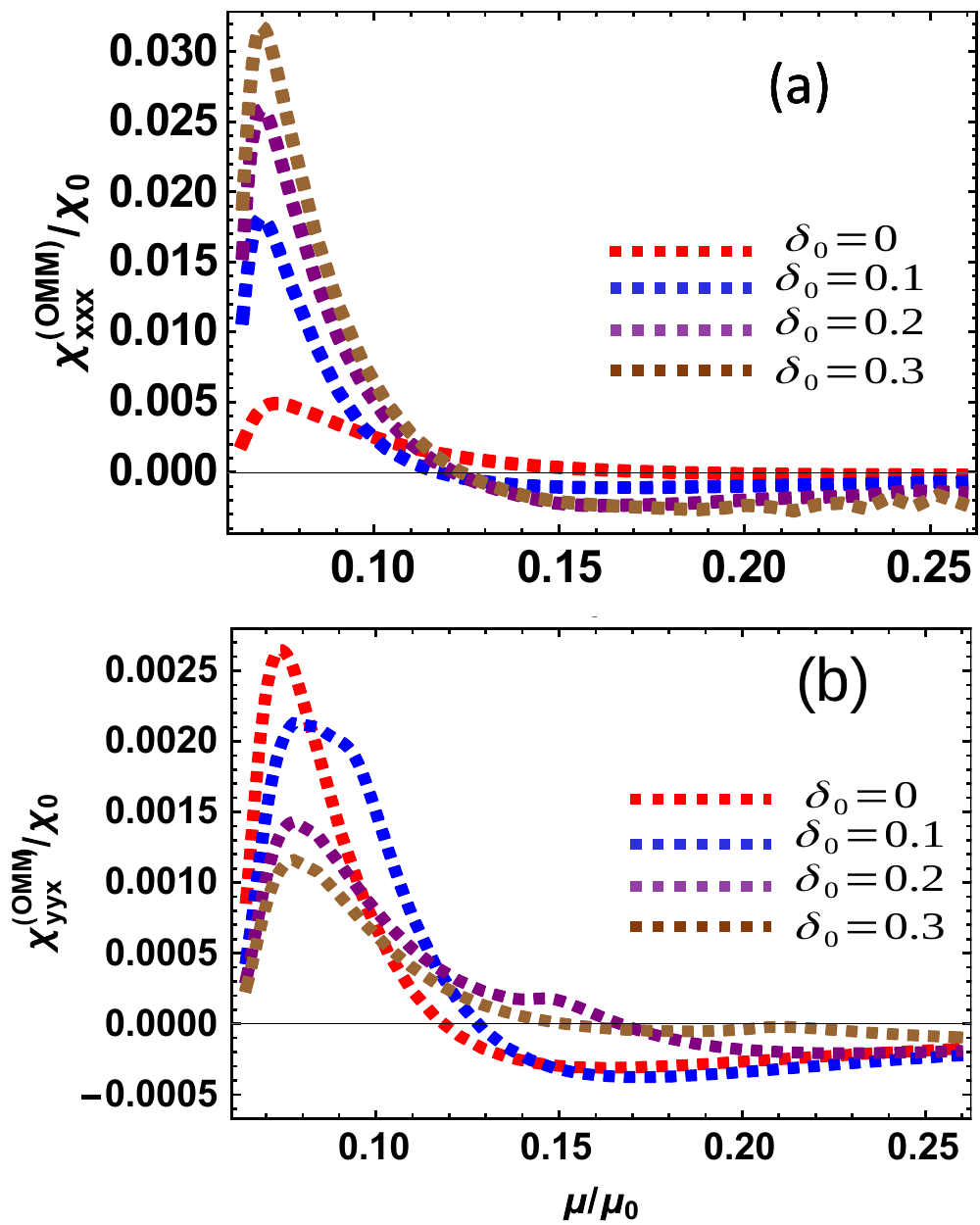}
\caption{(a)-(b) OMM induced SH conductivity as a function of Fermi energy for different values of $\delta_0$ in eV. The dashed color curves denote the results calculated numerically at $T=34$ K. The parameters used are the same as in Fig. \ref{fig2}.}
\label{fig5}
\end{figure}
We find that the $\sigma_{xy}^{\text{(OMM)}}$ varies as $\sqrt{\delta_0}$ in  $\gamma\rightarrow\infty$ limit. For low doping, $\sigma_{xy}^{\text{(OMM)}}$ decreases  with the Fermi energy whereas increases with $\delta_0$ . Above the saddle point, it continues to decrease with $\mu$ while increases with $\delta_{0}$. Note that this energy dependence of  $\sigma_{xy}^{\text{(OMM)}}$ is related to the fact that the magnitude of Berry curvature and OMM  decreases as the Fermi energy shifts away from the band edge. It is evident from Fig. \ref{fig2}(d) that the magnitude of OMM induced Hall conductivity is considerably smaller than the other linear contributions.
 
  We would like to point out that the linear conductivities $\sigma_{xx}^{\text{(D)}}$, $\sigma_{xy}^{\text{(L)}}$ and $\sigma_{xy}^{\text{(OMM)}}$ are predicted to be large for materials with small effective mass, whereas $\sigma_{yy}^{\text{(D)}}$ appears small for low effective mass. The effective mass of some proposed semi-Dirac materials are  $m^*= 13.6 m_e$ for (TiO$_2)_5/$(VO$_2)_3$, $m^*= 3.1 m_e$ ($\alpha$-(BEDT-TTF)$_2$I$_3$) and $m^*= 1.2*10^{-3}m_e$ (photonic crystals)\cite{kush}. We have used $m^*= 13.6 m_e$, $\beta =10^5$ m/s, $m_0=0.1$ eV, $\alpha$ = 2.7 meV$\cdot$nm$^2$, $B=2$ T and $\tau=10^{-12}$ s in this work. We also find that the linear conductivities decrease with the increase of gap parameter $m_0$ at a given value of $\mu$ and $\delta_0$. It is to be noted that only the Hall components of the above $B$-linear contribution to the conductivity survive due to the Onsager relation which implies $\sigma_{ij}(B)=\sigma_{ji}(-B)$. We have also calculated these contributions to the linear conductivities numerically at $T=34$ K and the obtained results at $T=34$ K match well with the analytical results evaluated at zero temperature.
  \begin{figure*}
    \centering
    \includegraphics[width=1.0\textwidth]{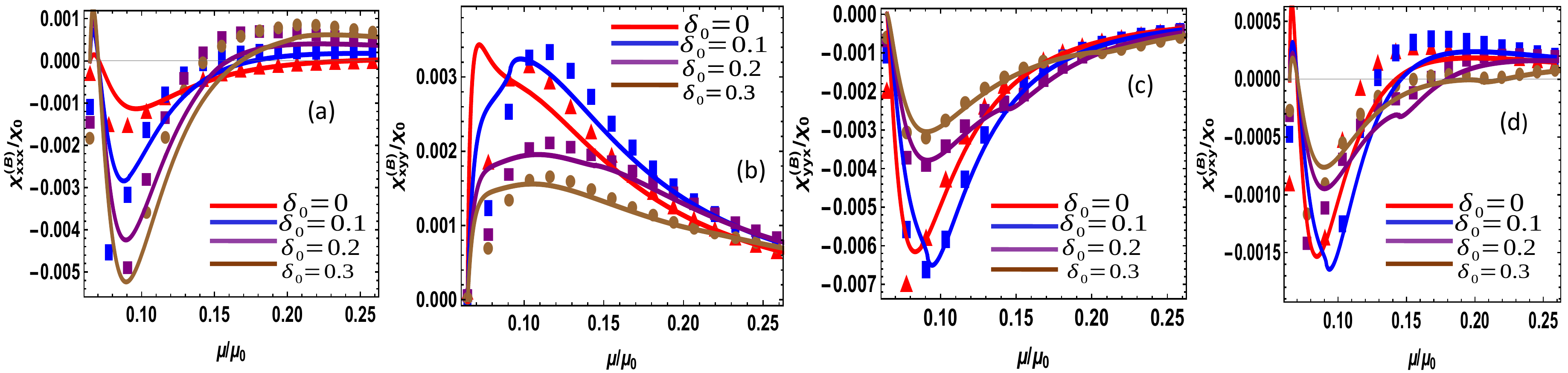}
    \caption{(a)-(d) Variation of different components of phase-space factor induced SH conductivity (measured in units of $\chi_0$) with the Fermi energy for different values of $\delta_0$ in eV. The solid color curves are the results calculated at zero temperature, while color plot markers represent the numerical results obtained at  $T=34$ K. The parameters used are the same as in Fig. \ref{fig2}.}
    \label{fig4}
\end{figure*} 
\begin{table*}
\centering

\setlength{\arrayrulewidth}{0.1mm}
\setlength{\tabcolsep}{2.5pt}
\renewcommand{\arraystretch}{0.8} 
  \begin{tabular}{|c|c|c|c|c|c|}
\hline

%& $B=0$& &$B\neq0$ \\ \cline{2-6}
%$j \propto {E^n}$& Drude &Anomalous Hall& Lorentz force and&phase-space factor&Orbital Magnetic \\
%&  & &anomalous velocity & (B)& moment(OMM)\\ \hline

\multirow{3}{*}{$j \propto {E^n}$} &
\multicolumn{2}{c|}{$B$=0 } &
 \multicolumn{3}{c|}{$B\neq0$ } \\ \cline{2-6}
 & Drude &Anomalous Hall& Lorentz force and&phase-space factor&Orbital Magnetic \\
 &  & &anomalous velocity & (B)& moment(OMM)\\ \hline
$n=1$& & & &  &\\
linear current responses& $\sigma_{xx}^{\text{(D)}}\neq\sigma_{yy}^{\text{(D)}}$&0 &  $\sigma_{xy}^{\text{(L)}}=-\sigma_{yx}^{\text{(L)}}$& 0& $\sigma_{xy}^{\text{(OMM)}}=-\sigma_{yx}^{\text{(OMM)}}$\\ \hline
& &  & & &\\
$n=2$& & & &$\chi_{xxx}^{\text{(B)}}$, $\chi_{xyy}^{\text{(B)}}$, &$\chi_{xxx}^{\text{(OMM)}}$, $\chi_{yxy}^{\text{(OMM)}}$=\\
SH current responses& 0&$\chi_{yxx}^{\text{(NAH)}}=-\chi_{xyx}^{\text{(NAH)}}$&$\chi_{xyy}^{\text{(NAL)}}=-\chi_{yxy}^{\text{(NAL)}}$&& \\
& &  &&$\chi_{yyx}^{\text{(B)}}, \chi_{yxy}^{\text{(B)}}$&$\chi_{xyy}^{\text{(OMM)}}=\chi_{yyx}^{\text{(OMM)}}$\\ \hline
\end{tabular}
 \caption{Highlighting the nonzero components of different contributions induced by the Berry curvature and the OMM to the linear and SH conductivities (up to linear order in a magnetic field).}
\label{table1}
\end{table*}
\subsection*{ Nonlinear conductivities}
\begin{center}
\textbf{A. Second-harmonic conductivities}
\end{center}

Next, we calculate the different contributions (arising from Berry curvature and OMM) to the SH conductivities of the system. Following the details of calculation similar to the linear case related to the integration over energy and $\phi$ in two regimes, the SH anomalous Hall conductivity can be evaluated using Eq. (\ref{NAH}) as 
\begin{equation}\label{nah}
\centering
\begin{aligned}
\chi_{xyx}^{\text{(NAH)}}&=-\frac{\chi_0 \tilde{m}_0}{6\sqrt{2}\pi^2 }\frac{({{\gamma}{\tilde{\delta}_0}})^{3/2}}{({{\tilde{\delta}_0^2}+\tilde{m}_0^2{\gamma}^{2}})^{3/2}}\\
 &\begin{cases}  \big[(1-\gamma) K(k)+2\gamma E(k)\big], 
 &\gamma < 1, \\ {2k}\big[(1-\gamma) K(k')+\gamma E(k')\big], 
 & \gamma \geq 1, \end{cases}
\end{aligned}
\end{equation}
where $\chi_0=e^3\tau\alpha/\hbar^3 \beta$. The other nonvanishing components of SH anomalous conductivity goes as $\chi_{yxx}^{\text{(NAH)}}=-\chi_{xyx}^{\text{(NAH)}}$. These off-diagonal terms are proportional to the $x$-component of the Berry curvature dipole which is nonzero due to the mirror symmetry along the $x$ axis\cite{kush}.

   We obtained the SH Hall conductivity arising due to the effective combination of Lorentz force and anomalous velocity for the system using Eq. (\ref{NL-lorentz}) and find it is related to SH anomalous Hall conductivity as
\begin{equation}\label{nal}
\chi_{yxy}^{\text{(NAL)}}=\frac{ \tilde{B}_1}{\mu/\mu_0}\chi_{xyx}^{\text{(NAH)}}.
\end{equation}
The other nonzero Hall component is  $\chi_{xyy}^{\text{(NAL)}}=-\chi_{yxy}^{\text{(NAL)}}$. It is evident from the above equation that the ratio $\chi_{yxy}^{\text{(NAL)}}/\chi_{xyx}^{\text{(NAH)}}$ is independent of $\delta_0$.
For the parameters used in our calculation and $\mu=0.2$ eV, we get ${ \tilde{B}_1}{\mu_0}/\mu=0.1$. Hence, $\chi_{yxy}^{\text{(NAL)}}$ is about an order of magnitude less than $\chi_{xyx}^{\text{(NAH)}}$.  Thus, the intrinsic anomalous response dominates over the Lorentz force contribution. Expanding the exact analytical results given by Eqs. (\ref{nah}) and (\ref{nal}) for $\gamma \geq 1$, we obtain
\begin{equation}
\begin{aligned}
\chi_{xyx}^{\text{(NAH)}}&\simeq\frac{-\chi_0 \tilde{m}_0}{64\pi }\frac{({{\gamma}{\tilde{\delta}_0}})^{3/2}}{({{\tilde{\delta}_0^2}+\tilde{m}_0^2{\gamma}^{2}})^{3/2}}\left[\frac{4\gamma+11}{{(1+\gamma)}^{3/2}}\right].\\
\chi_{yxy}^{\text{(NAL)}}&\simeq\frac{{\tilde{B}_1}{\gamma}}{\sqrt{{\tilde{\delta}_0^2}+{\tilde{m}_0^2}{\gamma}^{2}}}\chi_{xyx}^{\text{(NAH)}}.
\end{aligned}
\end{equation}
Near the band-edge at $\gamma\rightarrow \infty$, we find that $\chi_{xyx}^{\text{(NAH)}}$ and $\chi_{yxy}^{\text{(NAL)}} \propto {\mu^2}/{\sqrt{\delta_0}}$. For low doping, both the SH conductivities increase with $\mu$ but decrease with $\delta_0$. As Fermi energy is further increased, both $\chi_{xyx}^{\text{(NAH)}}$ and $\chi_{yxy}^{\text{(NAL)}}$ start decreasing with $\mu$, although a substantial change is not observed with $\delta_0$ in the region of high Fermi energy. A small kink is observed at the saddle point which reflects the change in Fermi surface topology. The Fermi energy dependence of $\chi_{xyx}^{\text{(NAH)}}$  and $\chi_{yxy}^{\text{(NAL)}}$ for different values of $\delta_0$ is depicted in Figs. \ref{fig3}(a) and \ref{fig3}(b) respectively. The peaks in $\chi_{yxy}^{\text{(NAL)}}$ appear to be  more pronounced than the peaks seen in $\chi_{xyx}^{\text{(NAH)}}$.

   We next turn to evaluate the OMM contribution to the SH conductivity using Eq. (\ref{NL-OMM}). We start by performing the integral over energy analytically using the approximation $f'_{\textnormal{eq}}=-\delta(\epsilon_{\mathbf{k}}-\mu)$ in the limit of $T\rightarrow0$, where we find that the resulting expression encounters divergence, unlike the previous cases. Thus the zero temperature approximation of the Dirac-delta function does not capture the proper results here. Therefore we proceed to calculate the OMM contribution by computing the results numerically at finite temperature $T=34$ K to overcome the issue of divergence. Figure \ref{fig5} represents the variation of OMM induced  SH conductivities ( $\chi_{xxx}^{\text{(OMM)}}$ and $\chi_{yxy}^{\text{(OMM)}}=\chi_{xyy}^{\text{(OMM)}}=\chi_{yyx}^{\text{(OMM)}}$) with the Fermi energy. 

  The phase-space contribution to the SH conductivity can be calculated using Eq. (\ref{NL-B}). Its expression after integration over energy (at zero temperature) is cumbersome and therefore not presented over here. The nonzero components of phase-space induced conductivity includes diagonal term  $\chi_{xxx}^{\text{(B)}}$ and off-diagonal components $\chi_{yyx}^{\text{(B)}}$, $\chi_{yxy}^{\text{(B)}}$  and $\sigma_{xyy}^{\text{(B)}}$. We have plotted the Fermi energy dependence of these terms for different values of $\delta_0$ in Fig. \ref{fig4}. Similar to the linear case, we have plotted the above three contributions to the SH conductivity by performing numerical calculations at temperature $T=34$ K and we observed that they agree closely with our analytical results obtained at zero temperature. Both the SH conductivities, $\chi_{abc}^{\text{(OMM)}}$ and $\chi_{abc}^{\text{(B)}}$ initially show an increase with the Fermi energy but then start decreasing. 
  
  We emphasize that all the above SH conductivities have a peak near the band edge which is related to the fact that these SH contributions arise from the Berry curvature and OMM and their magnitude decreases as the Fermi energy shifts from the band edge. The peaks are not observed exactly at the band edge because the SH contributions arise from the effective contribution of geometric quantities and band dispersion anisotropy. The nonzero components of these different contributions are highlighted in Table \ref{table1}. It is worth pointing out that all four contributions to the SH conductivities increase with the decrease in the effective band mass of semi-Dirac materials. Interestingly, these SH conductivities in a given semi-Dirac material are found to be comparable or order smaller than the SH conductivities of a 2D system hosting massive tilted Dirac fermions\cite{kamal}. We also noticed the variation of SH conductivities with the gap parameter $m_0$ and find that their peak is shifted with the increase of $m_0$. 
 \begin{center}
\textbf{B. Nonlinear DC current}
\end{center}
   We next proceed to calculate the NL dc current arising from these different contributions in the gapped semi-Dirac system  using Eqs. (\ref{dc}) and (\ref{2})-(\ref{B}),
\begin{widetext}
\begin{align}
{\mathbf{j}}^{(0)}_{\text{NAH}}&= \frac{\chi_{xyx,0}^{\text{(NAH)}}}{1+\omega^2\tau^2}\left[-2|E_x|^{2}\hat{\bf y}+{[E_{y}E_x^{*}]}_{+}\hat{\bf x}-i\omega\tau{[E_{y}E_x^{*}]}_{-}\hat{\bf x}\right],\\
{\mathbf{j}}^{(0)}_{\text{NAL}}&= \frac{\chi_{yxy,0}^{\text{(NAL)}}}{{(1+\omega^2\tau^2)}^2}\left[(1-\omega^2\tau^2)\left(-2|E_y|^{2}\hat{\bf x}+{[E_{y}E_x^{*}]}_{+}\hat{\bf y}\right)+2i\omega\tau{[E_{y}E_x^{*}]}_{-}\hat{\bf y}\right],\\
{\mathbf{j}}^{(0)}_{\text{OMM}}&= \frac{2}{1+{\omega^2}{\tau^2}}\Big[{\chi}_{xxx,0}^{\text{(OMM)}}{|E_x|}^{2}\hat{\bf x}+{\chi}_{yyx,0}^{\text{(OMM)}}\left({|E_y|}^{2}\hat{\bf x}+{[E_{y}E_x^{*}]}_{+}\hat{\bf y}\right)\Big],\\
{\mathbf{j}}^{(0)}_{\text{B}}&= \frac{1}{1+{\omega^2}{\tau^2}}\Big[2\left({\chi}_{xxx,0}^{\text{(B)}}{|E_x|}^{2}+{\chi}_{xyy,0}^{\text{(B)}}{|E_y|}^{2}\right)\hat{\bf x}+\left({\chi}_{yyx,0}^{\text{(B)}}+{\chi}_{yxy,0}^{\text{(B)}}\right){[E_{y}E_x^{*}]}_{+}\hat{\bf y}-i\omega\tau\left({\chi}_{yyx,0}^{\text{(B)}}-{\chi}_{yxy,0}^{\text{(B)}}\right){[E_{y}E_x^{*}]}_{-}\hat{\bf y} \Big].
\end{align}
\end{widetext}
 
 It is evident that the obtained NL dc current is dependent on the polarization of the incident electromagnetic wave. Here, we define the  total NL dc current as ${\mathbf{j}}^{(0)}_{\text{net}}={{j}}^{(0)}_{x}\hat{\bf x}+{{j}}^{(0)}_{y}\hat{\bf y}={\mathbf{j}}^{(0)}_{\text{NAH}}+{\mathbf{j}}^{(0)}_{\text{NAL}}+{\mathbf{j}}^{(0)}_{\text{OMM}}+{\mathbf{j}}^{(0)}_{\text{B}}$ and discuss their contributions in response to the linearly and circularly polarized light.\\
  \begin{figure*}
\centering
\includegraphics[width=1.0\textwidth]{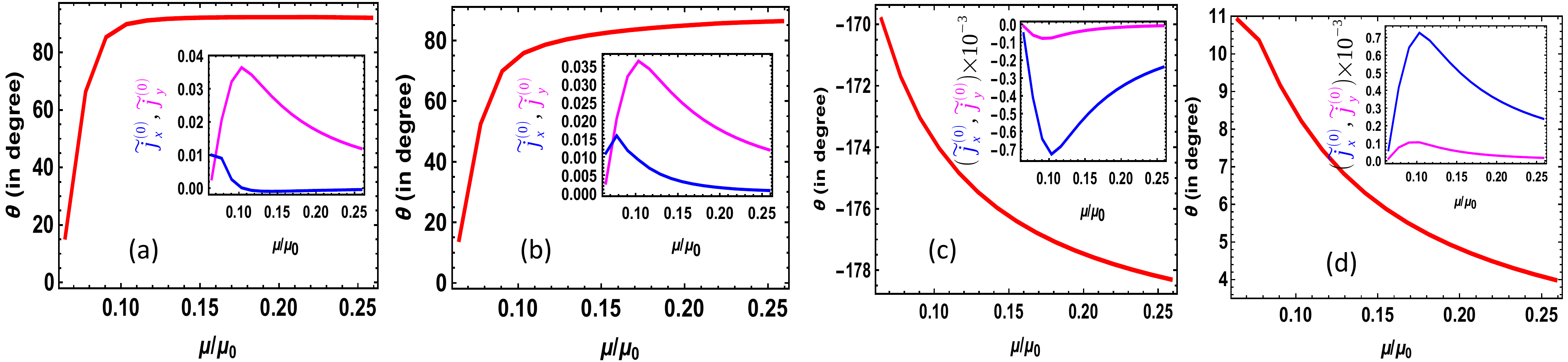}
\caption{ Variation of angle between the total NL dc current ${\mathbf{j}}^{(0)}_{\text{net}}$ and the $x$ axis (defined as $\theta$) with the Fermi energy  at $\delta_0=0.1$ eV for: (a) linearly polarized light along the $x$ direction, (b) left and right circularly polarized light in the limit of $\omega\tau \ll 1$, (c)-(d) left and right circularly polarized light in the $\omega\tau \gg 1$ limit. The corresponding insets represent the Fermi energy dependence of the total NL dc current in the $x$ and $y$ directions denoted by ${{j}}^{(0)}_{x}$ and ${{j}}^{(0)}_{y}$  respectively. We have defined the scaled dc current as $\tilde{j}_{x}^{(0)}= {j}_{x}^{(0)}/(2\chi_0|E_0|^2)$  and $\tilde{j}_{y}^{(0)}= {j}_{y}^{(0)}/(2\chi_0|E_0|^2)$. The other parameters used are the same as in Fig. \ref{fig2}.}
    \label{direct}
\end{figure*} 
   \textbf{Case I}: For linearly polarized light in the $x$ direction, i.e., $\mathbf{E}=(E_0,0,0)$, the total NL dc current along the $x$ and $y$ direction can be calculated as 
\begin{equation}
\begin{aligned}
{{j}}^{(0)}_{x}&= \frac{2}{1+\omega^2\tau^2}({\chi}_{xxx,0}^{\text{(OMM)}}+{\chi}_{xxx,0}^{\text{(B)}})|E_0|^2,\\
{{j}}^{(0)}_{y}&= -\frac{2}{1+\omega^2\tau^2}{\chi}_{xyx,0}^{\text{(NAH)}}|E_0|^2.
\end{aligned}
\end{equation}
In the low-frequency limit $\omega\tau \ll 1$, the above equation reduces to ${{j}}^{(0)}_{x}\approx 2({\chi}_{xxx,0}^{\text{(OMM)}}+{\chi}_{xxx,0}^{\text{(B)}})|E_0|^2$ and ${{j}}^{(0)}_{y}\approx -2{\chi}_{xyx,0}^{\text{(NAH)}}|E_0|^2$. We have plotted the variation of 
${{j}}^{(0)}_{x}$ and ${{j}}^{(0)}_{y}$ with the Fermi energy for $\delta_0=0.1$ eV in inset of Fig. \ref{direct}(a). For very low Fermi energy below the saddle point, ${{j}}^{(0)}_{x}$ and ${{j}}^{(0)}_{y}$ are comparable and thus the angle which the total DC current ${\mathbf{j}}^{(0)}_{\text{net}}$ makes with the $x$ axis (say, $\theta$) increases monotonically with the Fermi energy. For high Fermi energy above the saddle point, ${{j}}^{(0)}_{x}$ becomes vanishingly small and nearly constant as compared to ${{j}}^{(0)}_{y}$. Thus ${{j}}^{(0)}_{y}$ dominates and the angle between the ${\mathbf{j}}^{(0)}_{\text{net}}$ and $x$ axis saturates, $\theta\rightarrow {90}^\circ$ implying  that the total dc current is nearly perpendicular to the applied electric field as shown in  Fig. \ref{direct}(a). This can be attributed to the fact that ${{j}}^{(0)}_{y}\propto {\chi}_{xyx,0}^{\text{(NAH)}}$, which is finite due to the $x$ component of Berry curvature dipole. This Berry curvature dipole contribution arises from the mirror symmetry along the $x$ axis in the system. However, in the absence of $B$, the NL dc current is purely along the $y$-direction for any Fermi energy. 
     In the high frequency limit $\omega\tau \gg 1$, ${{j}}^{(0)}_{x}$ and ${{j}}^{(0)}_{y}$ are simply reduced by a factor of $\omega^2\tau^2$. 
    
     \textbf{Case II}: For linearly polarized light in the $y$ direction, i.e., $\mathbf{E}=(0,E_0,0)$, the NL dc current contribution in the $x$ and $y$ direction can be obtained as 
\begin{equation}
\begin{aligned}
{{j}}^{(0)}_{x}&= \frac{2}{1+\omega^2\tau^2}\Big[-\left(\frac{1-\omega^2\tau^2}{1+\omega^2\tau^2}\right){\chi}_{yxy,0}^{\text{(NAL)}}+{\chi}_{yyx,0}^{\text{(OMM)}}\\
&+{\chi}_{xyy,0}^{\text{(B)}}\Big]|E_0|^2,\\
{{j}}^{(0)}_{y}&= 0.
\end{aligned}
\end{equation}
In the $\omega\tau \ll 1$ limit, we obtain ${{j}}^{(0)}_{x}\approx 2(-{\chi}_{yxy,0}^{\text{(NAL)}}+{\chi}_{yyx,0}^{\text{(OMM)}}+{\chi}_{xyy,0}^{\text{(B)}})|E_0|^2$ and ${{j}}^{(0)}_{y}= 0$ which indicates the direction of ${\mathbf{j}}^{(0)}_{\text{net}}$ is directed along the positive $x$ direction, i.e., $\theta=0$. When $\omega\tau \gg 1$, the current in the $x$ direction is given by  ${{j}}^{(0)}_{x}\approx( 2/\omega^2\tau^2)({\chi}_{yxy,0}^{\text{(NAL)}}+{\chi}_{yyx,0}^{\text{(OMM)}}+{\chi}_{xyy,0}^{\text{(B)}})|E_0|^2$. The values of ${{j}}^{(0)}_{x}$ becomes negative due to which ${\mathbf{j}}^{(0)}_{\text{net}}$ is oriented antiparallel to the $x$ axis, i.e., $\theta=180^\circ$. These currents vanish in the absence of a magnetic field.

\textbf{Case III}:  For circularly polarized light, the electric field is given by $\mathbf{E}=(E_0,{\Lambda}iE_0,0)$, where the index $\Lambda=\pm$ indicates left and right circularly polarized light, respectively. The total NL dc current in the $x$ and $y$ direction can be obtained as
\begin{widetext}
\begin{equation} 
\begin{aligned}
{{j}}^{(0)}_{x}&= \frac{2}{1+\omega^2\tau^2}\left[\Lambda\omega\tau{\chi}_{xyx,0}^{\text{(NAH)}}-\left(\frac{1-\omega^2\tau^2}{1+\omega^2\tau^2}\right){\chi}_{yxy,0}^{\text{(NAL)}}+{\chi}_{xxx,0}^{\text{(OMM)}}+{\chi}_{yyx,0}^{\text{(OMM)}}+{\chi}_{xxx,0}^{\text{(B)}}+{\chi}_{xyy,0}^{\text{(B)}}\right]|E_0|^2,\\
{{j}}^{(0)}_{y}&= \frac{2}{1+\omega^2\tau^2}\left[-{\chi}_{xyx,0}^{\text{(NAH)}}-\Lambda\left(\frac{2\omega\tau}{1+\omega^2\tau^2}\right){\chi}_{yxy,0}^{\text{(NAL)}}+\Lambda\omega\tau({\chi}_{yyx,0}^{\text{(B)}}-{\chi}_{yxy,0}^{\text{(B)}})\right]|E_0|^2.
\end{aligned}
\end{equation}
\end{widetext}
\textbf{\textit{(a). In the $\omega\tau \ll 1$ limit}}: The terms in the above equation with the coefficient $\omega\tau$ correspond to the CPGE. Note that the CPGE vanishes in this limit, which implies that there is no difference between the effects of left and right circularly polarized light. For low Fermi energy, ${{j}}^{(0)}_{x}$ and ${{j}}^{(0)}_{y}$ are comparable due to which the angle between ${\mathbf{j}}^{(0)}_{\text{net}}$ and the $x$ axis increases with the Fermi energy. In the high Fermi energy region, both ${{j}}^{(0)}_{x}$ and ${{j}}^{(0)}_{y}$ decrease with the Fermi energy. However, the major contribution to the ${\mathbf{j}}^{(0)}_{\text{net}}$ comes from  ${{j}}^{(0)}_{y}\propto {\chi}_{xyx,0}^{\text{(NAH)}}$. Hence, $\theta$ shows a gradual increase with the Fermi energy, approaching the value close to $90^\circ$ (pointing in the direction close to the $y$ axis) as shown in Fig. \ref{direct}(b). In the absence of $B$, the NL dc current is purely along the y-direction for any Fermi energy.

\textbf{\textit{(b). In the  $\omega\tau \gg 1$ limit}}: The NL dc current in the $x$ and $y$ direction yields the form, ${{j}}^{(0)}_{x}\propto(1/\omega\tau){\chi}_{xyx,0}^{\text{(NAH)}}$ and ${{j}}^{(0)}_{y}\propto(1/\omega\tau)({\chi}_{yyx,0}^{\text{(B)}}-{\chi}_{yxy,0}^{\text{(B)}})$, respectively. Interestingly, we find that the current ${{j}}^{(0)}_{x}$ approaches the intrinsic value i.e., independent of scattering time $\tau$. The dominant contribution comes from   ${{j}}^{(0)}_{x}\propto{\chi}_{xyx,0}^{\text{(NAH)}}$. For left circularly polarized light, the current ${\mathbf{j}}^{(0)}_{\text{net}}$ flows nearly antiparallel to the $x$ axis, $\theta\rightarrow-180^\circ$ [Fig. \ref{direct}(c)]. The NL dc current points exactly along the negative $x$-direction in the absence of $B$. While for right circularly polarized light, ${\mathbf{j}}^{(0)}_{\text{net}}$ flows in a direction almost parallel to the $x$ axis with $\theta\rightarrow0$ [Fig. \ref{direct}(d)]. The NL dc current is directed purely along the positive $x$-direction without $B$, for any Fermi energy.

   Therefore, we conclude that the direction of the NL DC current depends on the Fermi energy, magnetic field and the polarization of the electromagnetic wave. It is predominantly governed by the underlying mirror symmetry of the system.
\begin{figure*}
    \centering
    \includegraphics[width=1.0\textwidth]{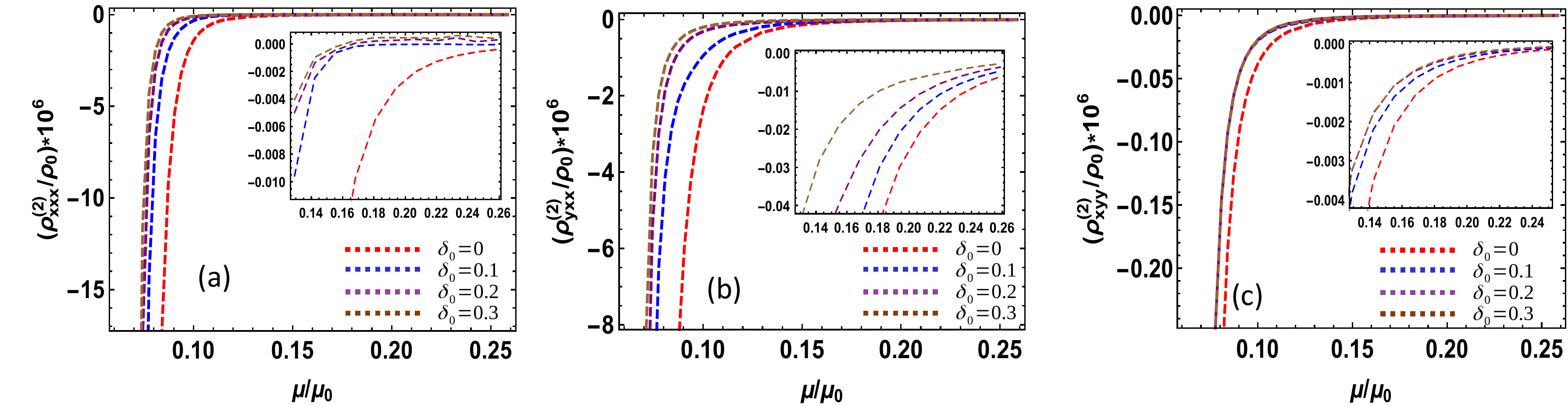}
    \caption{(a) SH resistivity ${\rho}_{xxx}^{(2)}$ and (b) SH Hall resistivity ${\rho}_{yxx}^{(2)}$ as a function of Fermi energy  for different values of $\delta_0$ (when current is applied along the $x$-direction). (c) The variation of SH Hall resistivity ${\rho}_{xyy}^{(2)}$ with the Fermi energy (when current is applied along the $y$-direction). The insets represent the blow-up of the region where Fermi energy $\mu$ ranges from 0.2 eV to 0.4 eV. The normalization parameter for SH resistivity is defined as $\rho_0=\alpha^4/(e^3\hbar^3\tau^2\beta^7)$. The dashed color curves represent the numerical results obtained at $T=34$ K. The parameters used are the same as in Fig. \ref{fig2}. }
    \label{fig6}
\end{figure*}
\section{Second-order nonlinear voltage responses}\label{V}
In the previous section, we have studied the linear, SH and NL dc current responses of the semi-Dirac system (through the conductivity tensors) on application of an oscillating electric field and a weak magnetic field. In this section, we look at the inverse process i.e. the induction of linear, SH and NL dc voltage/electric fields on passing an ac current through the system\cite{nl2,resis1,resis2,kamal}. Here, `NL' would imply the order of applied current.
% In response to an ac current $j^{\omega}$ flowing through the material in the experimental setups of NL transport\cite{nl2,resis1,resis2,kamal}, the induced linear electric field $E^{\omega}$, the induced NL DC electric field $E^{\text{nl-dc}}$ and the induced NL second harmonic electric field $E^{2\omega}$ are measured. 
These induced electric fields are related to the applied ac current $j^{\omega}$ by introducing resistivity into the picture. The linear resistivity $\rho_{ab}$ is defined as $E^{\omega}_{a}=\rho_{ab}j^{\omega}_{b}$, whereas the SH resistivity $\rho_{abc}^{(2\omega)}$ is given by $E^{2\omega}_{a}=\rho_{abc}^{(2\omega)}j^{\omega}_{b}j^{\omega}_{c}$ and the NL DC resistivity goes as $E^{\text{nl-dc}}_{a}=\rho_{abc}^{\text{nl-dc}}j^{\omega}_{b}j^{\omega}_{c}$. 
\subsection{Linear resistivity}
The linear resistivity matrix can be simply obtained by inverting the linear conductivity matrix. Keeping the lowest order magnetic field dependence, the linear resistivities can be calculated using $\rho_{xx}={1}/{\sigma_{xx}^{(D)}}$, $\rho_{xy}=-\rho_{yx}=-{\sigma_{xy}^{(L)}}/({{\sigma_{xx}^{(D)}\sigma_{yy}^{(D)}}})$ and $\rho_{yy}={1}/{\sigma_{yy}^{(D)}}$. Note that we have ignored the OMM induced Hall conductivity contribution as its magnitude is much small as compared to Lorentz force induced Hall conductivity, $\sigma_{xy}^{(\text{OMM})}\ll\sigma_{xy}^{\text{(L)}}$. We find that the linear longitudinal resistivity, $\rho_{xx}\propto B^{0}$ and linear Hall resistivity, $\rho_{xy}\propto B$. These linear resistivities show a decrease with Fermi energy. 
\subsection{Nonlinear resistivity}
Unlike the linear resistivity case, the  NL resistivity matrix cannot be obtained directly by inverting the NL conductivity matrix. Instead, the second-order NL resistivity is defined in terms of second-order NL conductivities and linear resistivities\cite{kamal} as
\begin{equation}\label{rho}
\rho_{abc}^{(2)}=-\rho_{ai}\chi_{ijk}\rho_{jb}\rho_{kc}.
\end{equation}
Note that Eq. (\ref{rho}) remains valid for both the SH resistivity and the NL dc resistivity, depending on which NL conductivity is employed on the right-hand side. Here, we present the results for the SH resistivity in the transport limit, $\omega\tau\ll 1$, obtained by applying the current along the $x$ and $y$ directions. It is worth pointing out that, in this limit, the SH resistivities are nearly equal to the NL dc resistivities. 

{\textbf{\textit{Case I. Current applied along the $x$-direction}}}: We consider that the current is flowing only along the $x$-direction, then the components of the SH resistivity matrix for 2D systems can be written as
\begin{equation}{\label{matrix1}}
\left( \begin{matrix}
{\rho}_{xxx}^{(2)} \\
{\rho}_{yxx}^{(2)} \\
\end{matrix}\right)
= -[\rho]
\left (\begin{matrix}
{\chi_{xxx}}& {\chi_{xxy}}&{\chi_{xyx}} &{\chi_{xyy}} \\
{\chi_{yxx}}& {\chi_{yxy}}&{\chi_{yyx}} &{\chi_{yyy}}
\end{matrix} \right)
\left( \begin{matrix}
{\rho}_{xx}^{2} \\
{\rho}_{xx}{\rho}_{yx} \\
{\rho}_{xx}{\rho}_{yx} \\
{\rho}_{yx}^{2}
\end{matrix}\right).
\end{equation}
Here, $\rho_{xxx}^{(2)}$ and $\rho_{yxx}^{(2)}$ denote the SH longitudinal  and SH Hall resistivities respectively and $[\rho]$ is the $2\times2$ linear resistivity matrix. We further simplify the above equation by keeping the lowest-order magnetic field dependence and ignoring the quadratic and higher $B$-field dependent terms lead to 
\begin{equation}
\begin{aligned}{\label{rho_xxx}}
{\rho}_{xxx}^{(2)}&=-\rho_{xx}^{2}\left({\chi_{xxx}}{\rho_{xx}}+{\chi_{xyx}}{\rho_{yx}}+{\chi_{yxx}}{\rho_{xy}}\right),\\
{\rho}_{yxx}^{(2)}&=-{\rho_{xx}^{2}}{\chi_{yxx}}{\rho_{yy}}.
\end{aligned}
\end{equation}
We find that the lowest-order magnetic field dependence of these SH resistivities goes as $\rho_{xxx}^{(2)}\propto B$ and $\rho_{yxx}^{(2)}\propto{B^{0}}$ which is distinct from the linear resistivities. It is evident from Eq. (\ref{rho_xxx}) that  $\rho_{xxx}^{(2)}$ depend on SH conductivities and linear Hall resistivities. Note that longitudinal resistivity  $\rho_{xxx}^{(2)}$ in the absence of magnetic field becomes zero as $\chi_{xxx}(B$=0)=0. Thus we emphasize that the predicted $B$-linear dependence of $\rho_{xxx}^{(2)}$ arises mainly from the diagonal component of SH conductivity ($\chi_{xxx}=\chi_{xxx}^{\text{(OMM)}}+\chi_{xxx}^{\text{(B)}}$) which remains finite due to the surviving mirror symmetry along the $x$-direction. The NL resistivities originate from Berry curvature and OMM and therefore has a quantum mechanical origin. The SH resistivities ${\rho}_{xxx}^{(2)}$ and ${\rho}_{yxx}^{(2)}$ show a relatively significant decrease with the Fermi energy for low doping as compared to high doping which is illustrated in Figs. \ref{fig6}(a) and \ref{fig6}(b). Furthermore, the scattering time dependence of SH resistivities is found to be  ${\rho}_{xxx}^{(2)}\propto 1/\tau$ and ${\rho}_{yxx}^{(2)}\propto 1/{\tau^2}$. Thus, the experimentally connected scattering time independent ratios are defined for SH  resistivities as ${\rho}_{xxx}^{(2)}/{{\rho}_{xx}}$ and ${\rho}_{yxx}^{(2)}/{{\rho}_{xx}^{2}}$. We find that ${\rho}_{xxx}^{(2)}/{{\rho}_{xx}}$ ($\sim$ ratio of  SH longitudinal voltage to the product of linear voltage and current) decreases with the Fermi energy and $\delta_0$  whereas ${\rho}_{yxx}^{(2)}/{{\rho}_{xx}^{2}}$ ($\sim$ ratio of SH Hall voltage to the square of linear voltage) decreases with Fermi energy but independent of $\delta_0$. Unlike the linear and SH conductivities, the change in Fermi surface topology is not reflected in the SH resistivities through the kink.\\

\begin{table}
\centering

\setlength{\arrayrulewidth}{0.1mm}
\setlength{\tabcolsep}{2.5pt}
\renewcommand{\arraystretch}{0.35} 
  \begin{tabular}{|c|c|c|}
\hline
%& $B=0$& &$B\neq0$ \\ \cline{2-3}
%$j \propto {E^n}$& Drude &Anomalous Hall& Lorentz force and&phase-space factor&Orbital Magnetic \\
%&  & &anomalous velocity & (B)& moment(OMM)\\ \hline
& &\\
$E \propto {j^n}$ &
$B$=0 &
 $B\neq0$ \\
 & & \\ \hline
& &\\
$n$=1& &  \\
& &\\
linear voltage responses& $\rho_{xx}$, $\rho_{yy}$&$\rho_{xy}$ = $-\rho_{yx}$ \\
& &\\ 
$E^{\omega}_{a}=\rho_{ab}j^{\omega}_{b}$& &   \\
& & \\ \hline
& & \\
$n=2$& &  \\
& & \\
SH voltage responses& $\rho_{yxx}^{(2)}$&$\rho_{xxx}^{(2)}$, $\rho_{xyy}^{(2)}$ \\
& &  \\ 
$E^{2\omega}_{a}=\rho_{abb}^{(2\omega)}{(j^{\omega}_{b})}^2$& &   \\
& & \\ \hline
\end{tabular}
 \caption{Highlighting the nonzero components of  linear and SH resistivities up to linear order in a magnetic field.}
\label{table-2}
\end{table}
{\textbf{\textit{Case II. Current applied along the $y$-direction}}}: In the wake of anisotropic energy dispersion, the current is also applied along the $y$-direction and the corresponding SH resistivity matrix turns out to be
\begin{equation}{\label{matrix1}}
\left( \begin{matrix}
{\rho}_{xyy}^{(2)} \\
{\rho}_{yyy}^{(2)} \\
\end{matrix}\right)
= -[\rho]
\left (\begin{matrix}
{\chi_{xxx}}& {\chi_{xxy}}&{\chi_{xyx}} &{\chi_{xyy}} \\
{\chi_{yxx}}& {\chi_{yxy}}&{\chi_{yyx}} &{\chi_{yyy}}
\end{matrix} \right)
\left( \begin{matrix}
{\rho}_{xy}^{2} \\
{\rho}_{xy}{\rho}_{yy} \\
{\rho}_{xy}{\rho}_{yy} \\
{\rho}_{yy}^{2}
\end{matrix}\right).
\end{equation}
Focusing on the lowest-order magnetic field dependence terms, we get
\begin{equation}
\begin{aligned}
{\rho}_{yyy}^{(2)}&=0,\\
{\rho}_{xyy}^{(2)}&=-{\rho_{xx}}{\rho_{yy}}({\chi_{xyx}}{\rho_{xy}}+{\chi_{xyy}}{\rho_{yy}}).
\end{aligned}
\end{equation}
Here, the SH longitudinal resistivity ${\rho}_{yyy}^{(2)}$ vanishes and the SH Hall resistivity ${\rho}_{xyy}^{(2)}\propto B$.   The SH Hall resistivity  $\rho_{xyy}^{(2)}$ in the absence of magnetic field comes zero as $\chi_{xyy}(B=0)=0$. Hence this $B$-linear dependence of ${\rho}_{xyy}^{(2)}$ originally arises from the SH Hall conductivity ($\chi_{xyy}=\chi_{xyy}^{(\text{NAL})}+\chi_{xyy}^{(\text{OMM})}+\chi_{xyy}^{(\text{B})}$) which elucidates that ${\rho}_{xyy}^{(2)}$ also depends on intrinsic band geometric quantities. The SH Hall resistivity ${\rho}_{xyy}^{(2)}$ decreases with the Fermi energy as shown in Fig. \ref{fig6}(c). The scattering time dependence goes as ${\rho}_{xyy}^{(2)}\propto 1/\tau$ and the ratio ${\rho}_{xyy}^{(2)}/\rho_{yy}\propto\tau^{0}$. We notice that the ratio ${\rho}_{xyy}^{(2)}/\rho_{yy}$ decreases with the Fermi energy and $\delta_0$. The nonzero components of resistivities are summarized in Table \ref{table-2}.

Some key observations about the SH resistivities of this system are as follows: A SH Hall resistivity can be observed even in the absence of $B$, when a current is applied along the $x$-direction. A SH longitudinal resistivity can be induced along the $x$-direction upon the application of $B$. This can also be seen as the induction of second-order longitudinal magnetoresistance in a system due to the effect of geometric quantities. On the other hand, for a current along the $y$-direction, a SH Hall resistivity can be brought about upon the application of $B$. The SH longitudinal resistivity along the $y$-direction is zero, irrespective of the presence or absence of $B$ (up to linear order in $B$). These observations highlight the intriguing asymmetry in the dependence on the magnetic field between the two directions. It is clear that the SH resistivities of this system are influenced by both geometric quantities and the underlying mirror symmetry, contributing to the distinct responses observed along the $x$ and $y$ directions.
\section{conclusions}\label{VI}
In this study, we investigated the linear and second-order NL current and voltage responses of a 2D gapped semi-Dirac system with merging Dirac nodes. We analyzed the system's response to a weak magnetic field, using the semiclassical Boltzmann formalism. Notably, the system possesses an intrinsic NL anomalous Hall conductivity, which can be attributed to the underlying mirror symmetry\cite{kush}. The application of a magnetic field introduces three distinct contributions to the SH conductivity tensor, originating from geometric quantities: the combined effects of anomalous velocity and Lorentz force, the OMM, and the Berry curvature correction to the phase-space factor. 

    We have obtained exact analytical expressions for the  linear and SH conductivities of the system, expressed  in terms of complete elliptic integrals of the first and second kind, within the transport limit. Our findings have facilitated a comprehensive analysis of the conductivities  and their dependence on Fermi energy and $\delta_0$. For $\delta_0 > 0$, the change in Fermi surface topology (from a single-connected Fermi surface for high Fermi energy to two Fermi surfaces for low Fermi energy) is also manifested in the behavior of both the obtained linear and SH conductivities.  A small kink is observed at the saddle point $\gamma=1$ for the fixed positive values of $\delta_0$ which marks the transition between two different type of Fermi surfaces. We obtained approximate expressions for conductivities at low Fermi energy and explicitly showed their dependence on Fermi energy and $\delta_0$ near the band edge. We found that the geometric mean (ratio) of Drude conductivities in the $x$ and $y$ directions is independent of $\delta_0$ for low (high) doping. We showed that the ratio of the anomalous velocity and Lorentz force-induced SH Hall conductivity to the SH anomalous Hall conductivity is independent of $\delta_0$ and inversely related to Fermi energy. It is to be noted that in a time-reversal symmetric system, SH Drude conductivity vanishes which implies that the SH longitudinal magnetoconductivity arises solely due to geometric quantities. 

       We have also examined the zero-frequency current response generated at the second-order in the electric field. We have observed that the orientation of the NL dc current depends on the Fermi energy, magnetic field and polarization of electromagnetic wave. In the absence of magnetic field, the NL dc currents are oriented purely along the $y$-direction for $x$-polarized and low-frequency circularly polarized light ($\omega\tau \ll 1$), whereas they align along the $x$-direction for high-frequency circularly polarized light ($\omega\tau \gg 1$). It is worth noting that the NL dc current vanishes entirely for $y$-polarized light in absence of $B$. These orientations are predominantly governed by the underlying mirror symmetry of the system. In the presence of a magnetic field, the Fermi energy of the system serves as a control parameter for rotating the net NL dc current vector. Substantial rotations can be achieved by applying $x$-polarized and low-frequency circularly polarized light ($\omega\tau \ll 1$) compared to high-frequency circular polarization ($\omega\tau \gg 1$). However, no rotation is observed for $y$-polarized light. For $y$-polarized light, the NL dc current is consistently parallel or antiparallel to the $x$-axis. However, for $x$-polarized light, the NL DC current is nearly $y$-directed at high Fermi energies. In the high frequency limit, for circular polarization, the NL dc current  aligns in a direction close to $x$-axis. Conversely, at low frequencies, the CPGE current vanishes  and the NL dc current is directed close to the $y$-axis for high fermi energies. This conversion of applied ac electric field into DC current (rectification) in proposed semi-Dirac materials may hold applications for a wide range of technologies\cite{isobe}.

  We further study the SH magnetoresistivities of the system for two different orientations of current flow: (1) along $x$ and (2) along $y$ directions. For the current applied along the $x$-direction, the SH longitudinal resistivity scales linearly with the magnetic field, while SH Hall resistivity is independent of $B$ in the lowest $B$-order. However, for the current applied along the $y$-direction, the SH longitudinal resistivity vanishes in both zeroth and linear order of $B$ and SH Hall resistivity varies linearly with $B$ in the lowest order. The predicted $B$-linear dependence of SH resistivities mainly arises from SH conductivities, which stem from band geometric quantities and underlying mirror symmetry of the system. Our results provide the platform for understanding the second-order NL  magnetotransport induced by geometric quantities in an anisotropic 2D system undergoing a topological transition with merging of two Dirac points.
\begin{center}
	{\bf ACKNOWLEDGEMENTS}
\end{center}
We would like to thank Kamal Das for
useful discussions.

\appendix{}
\section{Linear current responses}\label{a}
In this appendix, we calculate the current responses linear in $E$ and up to linear order in $B$. To obtain the NDF linear in $E$, we use the ansatz $f_1(t)=f_{1}^{\omega}e^{-i{\omega}t}+f_{1}^{\omega\ast}e^{i{\omega}t}$ in Eq. (\ref{BTE}) and obtain
%and equating the terms corresponding to $e^{-i{\omega}t}$ leads to
%\begin{equation}
%-i\omega f_{1}^{\omega}-\frac{e}{\hbar D}\Big[\frac{1}{2}(\mathbf{E}\cdot\boldsymbol{\nabla}_{\mathbf{k}}\tilde{f}_{\textnormal{eq}})+(\tilde{\boldsymbol{v}}_{\mathbf{k}} \times \mathbf{B})\cdot\boldsymbol{\nabla}_{\mathbf{k}}f_{1}^{\omega}\Big]=-\frac{ f_{1}^{\omega}}{\tau}.
%\end{equation}
\begin{equation}
f_{1}^{\omega}=\frac{1}{2}\sum_{\eta=0}^{\infty}\left(\frac{{\tau_\omega}{\hat{L}_{B}}}{D}\right)^{\eta}\left(\frac{e\tau_{\omega}}{\hbar{D}}\mathbf{E}\cdot\boldsymbol{\nabla}_{\mathbf{k}}\tilde{f}_{\textnormal{eq}}\right),
\end{equation}
where $\tau_{\omega}=\tau/(1-i\omega\tau)$. The current responses linear in $E$ can be expressed as $\textbf{j}_{1}{(t)}=\textbf{j}_{10}(t)+\textbf{j}_{11}(t)$. The magnetic field independent current of fundamental frequency can be expressed as $\textbf{j}_{10}(t)=\textbf{j}_{10}^{\omega}e^{-i{\omega}t}+\textbf{j}_{10}^{\omega\ast}e^{i{\omega}t}$, where we obtain
\begin{equation}
\textbf{j}_{10}^{\omega}=-\frac{e^2}{2\hbar}\int[d\mathbf{k}]\left[{\tau}_{\omega}{\boldsymbol{v}}_{\mathbf{k}}(\mathbf{E}\cdot\boldsymbol{\nabla}_{\mathbf{k}}){f}_{\textnormal{eq}}+(\mathbf{E} \times  \boldsymbol{\Omega}){f}_{\textnormal{eq}}\right].
\end{equation}
The magnetic field independent conductivities take the following form
\begin{align}
\label{drude}
\sigma_{ab}^{\text{(D)}}&=-\frac{e^2 \tau_{\omega}}{2}\int[d\mathbf{k}]v_a v_b f'_{\textnormal{eq}}.\\
\label{AHC}
\sigma_{ab}^{\text{(AHC)}}&=-\frac{e^2}{2\hbar} \varepsilon_{abd}\int[d\mathbf{k}]\Omega_d f_{\textnormal{eq}} .
\end{align}
Equation (\ref{drude}) refers to the Drude conductivity, while Eq. (\ref{AHC}) describes the intrinsic anomalous
Hall conductivity which is independent of scattering time and vanishes for the TRS preserved system.

  The magnetic field dependent current of fundamental frequency can be written as $\textbf{j}_{11}(t)=\textbf{j}_{11}^{\omega}e^{-i{\omega}t}+\textbf{j}_{11}^{\omega\ast}e^{i{\omega}t}$, we get
\begin{equation}\label{j11}
\centering
\begin{aligned}
\textbf{j}_{11}^{\omega}& =\frac{e^2}{2\hbar}\int[d\mathbf{k}]\bigg[(\mathbf{E} \times  \boldsymbol{\Omega}){\epsilon_m} {f'_{\textnormal{eq}}}+ {\tau_{\omega}}{\boldsymbol{v}}_{\mathbf{k}}\big\{\frac{e}{\hbar}(\boldsymbol{\Omega}\cdot\mathbf{B})\\
&(\mathbf{E}\cdot\boldsymbol{\nabla}_{\mathbf{k}}){f}_{\textnormal{eq}}+(\mathbf{E}\cdot\boldsymbol{\nabla}_{\mathbf{k}}){\epsilon_m} {f'_{\textnormal{eq}}}\big\}-{\tau}_{\omega}^{2}{\boldsymbol{v}}_{\mathbf{k}}\hat{L}(\mathbf{E}\cdot \boldsymbol{\nabla}_{\mathbf{k}}){f}_{\textnormal{eq}}\bigg],
\end{aligned}
\end{equation}
where $\hat{L}=({e}/{\hbar})[({\boldsymbol{v}}_{\mathbf{k}} \times \mathbf{B})\cdot\boldsymbol{\nabla}_{\mathbf{k}}]$. In the above expression, the terms  proportional to $\tau$ vanish since $\Omega_{k}$, ${v_k}$ and $\epsilon_m$ are odd in the presence of TRS. The magnetic field-dependent conductivities which survive under TRS are given by
\begin{align}
\label{lorentz}
\sigma_{ab}^{\text{(L)}}&=-\frac{e^3 \tau_{\omega}^{2}B}{2\hbar}\int[d\mathbf{k}]v_a\left( v_y \partial_{{k}_{x}} - v_x  \partial_{{k}_{y}}\right){v_b}f'_{\textnormal{eq}}.\\
\label{omm}
\sigma_{ab}^{\text{(OMM)}}&=\frac{e^2}{2\hbar}\varepsilon_{abd}\int[d\mathbf{k}]\Omega_d \epsilon_m f'_{\textnormal{eq}}.
\end{align}
Equation (\ref{lorentz}) represents the Lorentz force contribution (classical Hall effect), while Eq. (\ref{omm}) describes the OMM-induced Hall effect.
\section{Nonlinear DC current}\label{b}
In this Appendix, we provide the general expressions for the various contributions induced by Berry curvature and OMM to the NL DC current in the presence of magnetic field. These expressions are valid for all 2D systems and are  presented in terms of the polarization of the electromagnetic wave.
 
   The NL dc current emerging from the interplay of anomalous velocity and Lorentz force can be obtained as
%\begin{equation}
%\centering
%\begin{aligned}
%{\mathbf{j}}^{(0)}_{\text{NAL}}&= \frac{1}{(1+{\omega^2}{\tau^2})^2}\Big[(1-{\omega^2}{\tau^2})\\
%&\Big\{2\left( {\chi}_{xyy}^{\text{(NAL)}}{|E_y|}^{2}\hat{\bf x}
%+{\chi}_{yxx}^{\text{(NAL)}}{|E_x|}^{2}\hat{\bf y} \right)\\
%&+\left({\chi}_{xyx}^{\text{(NAL)}}{[E_{y}E_x^{*}]}_{+}\hat{\bf x}+{\chi}_{yxy}^{\text{(NAL)}}{[E_{y}E_{x}^{*}]}_{+}\hat{\bf y}%\right)\Big\}\\
%& - 2i\omega\tau\left({\chi}_{xyx}^{\text{(NAL)}}{[E_{y}E_x^{*}]}_{-}\hat{\bf x}-{\chi}_{yxy}^{\text{(NAL)}}{[E_{y}E_x^{*}]}%_{-}\hat{\bf y}\right)\Big].
%\end{aligned}
%\end{equation}
\begin{widetext}
\begin{equation}\label{2}
\centering
\begin{aligned}
{\mathbf{j}}^{(0)}_{\text{NAL}}&= \frac{1}{(1+{\omega^2}{\tau^2})^2}\Big[(1-{\omega^2}{\tau^2})
\Big\{2\left( {\chi}_{xyy,0}^{\text{(NAL)}}{|E_y|}^{2}\hat{\bf x}
+{\chi}_{yxx,0}^{\text{(NAL)}}{|E_x|}^{2}\hat{\bf y} \right)+\left({\chi}_{xyx,0}^{\text{(NAL)}}{[E_{y}E_x^{*}]}_{+}\hat{\bf x}+{\chi}_{yxy,0}^{\text{(NAL)}}{[E_{y}E_{x}^{*}]}_{+}\hat{\bf y}\right)\Big\}\\
& - 2i\omega\tau\left({\chi}_{xyx,0}^{\text{(NAL)}}{[E_{y}E_x^{*}]}_{-}\hat{\bf x}-{\chi}_{yxy,0}^{\text{(NAL)}}{[E_{y}E_x^{*}]}_{-}\hat{\bf y}\right)\Big].
\end{aligned}
\end{equation}
The OMM contribution to the NL dc current can be calculated as
\begin{equation}\label{3}
\begin{aligned}
{\mathbf{j}}^{(0)}_{\text{OMM}}&= \frac{1}{1+{\omega^2}{\tau^2}}\Big[2\left({\chi}_{xxx,0}^{\text{(OMM)}}{|E_x|}^{2}+{\chi}_{xyy,0}^{\text{(OMM)}}{|E_y|}^{2}\right)\hat{\bf x}+2\left({\chi}_{yxx,0}^{\text{(OMM)}}{|E_x|}^{2}+{\chi}_{yyy,0}^{\text{(OMM)}}{|E_y|}^{2}\right)\hat{\bf y}\\
&+\left({\chi}_{xxy,0}^{\text{(OMM)}}+{\chi}_{xyx,0}^{\text{(OMM)}}
 \right){[E_{y}E_x^{*}]}_{+}\hat{\bf x}+\left({\chi}_{yxy,0}^{\text{(OMM)}}+{\chi}_{yyx,0}^{\text{(OMM)}}
 \right){[E_{y}E_x^{*}]}_{+}\hat{\bf y}  \\
 &-i\omega\tau\left\{\Big({\chi}_{xyx,0}^{\text{(OMM)}}-{\chi}_{xxy,0}^{\text{(OMM)}}
 \right){[E_{y}E_x^{*}]}_{-}\hat{\bf x}+
\left({\chi}_{yyx,0}^{\text{(OMM)}}-{\chi}_{yxy,0}^{\text{(OMM)}}
 \right){[E_{y}E_x^{*}]}_{-}\hat{\bf y}\Big\} \Big],
\end{aligned}
\end{equation}
\end{widetext}
The NL dc current induced by the phase-space factor can be obtained as
\begin{equation}\label{B}
{\mathbf{j}}^{(0)}_{\text{B}}={\mathbf{j}}^{(0)}_{\text{OMM}}\left(\chi_{abc,0}^{\text{(OMM)}}\rightarrow\chi_{abc,0}^{\text{(B)}}\right),
\end{equation}
where $\chi_{abc,0}^{\text{(NAL)}}=\chi_{abc}^{\text{(NAL)}}(\omega=0)$, $\chi_{abc,0}^{\text{(OMM)}}=\chi_{abc}^{\text{(OMM)}}(\omega=0)$ and $\chi_{abc,0}^{\text{(B)}}=\chi_{abc}^{\text{(B)}}(\omega=0)$ are the corresponding SH conductivities given by Eqs. (\ref{NL-lorentz})-(\ref{NL-B}), respectively, for $\omega= 0$.

\end{document}